\documentclass[conference]{IEEEtran}
\IEEEoverridecommandlockouts
\usepackage{cite}
\usepackage{amsmath,amssymb,amsfonts}
\usepackage{algorithmic}
\usepackage{graphicx}
\usepackage{textcomp}
\usepackage{xcolor}
\usepackage{url}

\ifCLASSOPTIONcompsoc
    \usepackage[caption=false, font=normalsize, labelfont=sf, textfont=sf]{subfig}
\else
    \usepackage[caption=false, font=footnotesize]{subfig}
\fi

\usepackage{tikz}
\usepackage{tabularx}
\usepackage{booktabs} 
\usepackage{multirow}
\usepackage[ruled,vlined]{algorithm2e}  

\usepackage{listings}
\makeatletter
\newcommand{\removelatexerror}{\let\@latex@error\@gobble}
\makeatother

\def\BibTeX{{\rm B\kern-.05em{\sc i\kern-.025em b}\kern-.08em
    T\kern-.1667em\lower.7ex\hbox{E}\kern-.125emX}}
\begin{document}

\pagestyle{empty}
\onecolumn

{\Large \bfseries Clustered Federated Learning Architecture for Network Anomaly Detection in Large Scale Heterogeneous IoT Networks}

\vspace{3cm}

{\LARGE This article has been accepted for publication in Computers \& Security}

\vspace{1cm}

{\LARGE Citation information: \url{https://doi.org/10.1016/j.cose.2023.103299}}

\vspace{3cm}

{\Large This is a PDF file of an article that has undergone enhancements after acceptance, such as the addition of a cover page and metadata, and formatting for readability, but it is not yet the definitive version of record. This version will undergo additional copyediting, typesetting and review before it is published in its final form, but we are providing this version to give early visibility of the article. Please note that, during the production process, errors may be discovered which could affect the content, and all legal disclaimers that apply to the journal pertain.}

\twocolumn

\clearpage
\pagenumbering{arabic}

\pagestyle{plain} 

\title{Clustered Federated Learning Architecture for Network Anomaly Detection in Large Scale Heterogeneous IoT Networks
\thanks{}
}

\author{
\IEEEauthorblockN{Xabier Sáez-de-Cámara\IEEEauthorrefmark{1}\IEEEauthorrefmark{2}\IEEEauthorrefmark{3}, Jose Luis Flores\IEEEauthorrefmark{1}, Cristóbal Arellano\IEEEauthorrefmark{1}, Aitor Urbieta\IEEEauthorrefmark{1} and Urko Zurutuza\IEEEauthorrefmark{2}}
\IEEEauthorblockA{\IEEEauthorrefmark{1}Ikerlan Technology Research Centre, Basque Research and Technology Alliance (BRTA)\\Arrasate-Mondragón, Spain\\\{xsaezdecamara, jlflores, carellano, aurbieta\}@ikerlan.es}\IEEEauthorblockA{\IEEEauthorrefmark{2}Mondragon Unibertsitatea\\Arrasate-Mondragón, Spain\\uzurutuza@mondragon.edu}\IEEEauthorblockA{\IEEEauthorrefmark{3}Corresponding author}
}

\maketitle

\begin{abstract}

There is a growing trend of cyberattacks against Internet of Things (IoT) devices; moreover, the sophistication and motivation of those attacks is increasing. The vast scale of IoT, diverse hardware and software, and being typically placed in uncontrolled environments make traditional IT security mechanisms such as signature-based intrusion detection and prevention systems challenging to integrate. They also struggle to cope with the rapidly evolving IoT threat landscape due to long delays between the analysis and publication of the detection rules. Machine learning methods have shown faster response to emerging threats; however, model training architectures like cloud or edge computing face multiple drawbacks in IoT settings, including network overhead and data isolation arising from the large scale and heterogeneity that characterizes these networks.

This work presents an architecture for training unsupervised models for network intrusion detection in large, distributed IoT and Industrial IoT (IIoT) deployments. We leverage Federated Learning (FL) to collaboratively train between peers and reduce isolation and network overhead problems. We build upon it to include an unsupervised device clustering algorithm fully integrated into the FL pipeline to address the heterogeneity issues that arise in FL settings. The architecture is implemented and evaluated using a testbed that includes various emulated IoT/IIoT devices and attackers interacting in a complex network topology comprising 100 emulated devices, 30 switches and 10 routers. The anomaly detection models are evaluated on real attacks performed by the testbed's threat actors, including the entire Mirai malware lifecycle, an additional botnet based on the Merlin command and control server and other red-teaming tools performing scanning activities and multiple attacks targeting the emulated devices.

\end{abstract}

\begin{IEEEkeywords}
Anomaly detection, Botnet, Internet of Things, Intrusion detection, Machine learning, Network security
\end{IEEEkeywords}

\section{Introduction}

The ever-growing adoption of the Internet of Things (IoT) is enabling manufacturers multiple benefits, such as productivity boosts, increased automation, cost savings, and the minimization of production errors and waste. This is achieved by connecting Internet-enabled devices, Cyber-Physical Systems (CPS) and other ``things'' with the ability to collect, monitor, analyze and share data to make decisions and interact with physical processes, often with little or no human intervention~\cite{Boyes2018}. However, this high level of connectivity also brings a higher risk of cybersecurity breaches and a bigger attack surface both for domestic and industrial IoT devices~\cite{Sisinni2018, Rehman2019}. Especially nowadays that many solutions are being replaced with commercial off-the-shelf devices~\cite{Boyes2018} that prioritize fast market adoption and convenience over security~\cite{Jalali2019}.

Poor security practices and vulnerabilities~\cite{Neshenko2019}, coupled with the mass adoption and high interconnectivity, make IoT an attractive target for malware designers. A notable example is the 2016~Mirai worm~\cite{Antonakakis2017}, which exploited the widespread use of weak or hardcoded passwords to compromise a diverse set of devices from various manufacturers. The sophistication of those threats continues to grow. Newer Mirai variants developed after the public release of its source code~\cite{Miraisrc} and other advanced IoT malware such as Mozi\footnote{https://malpedia.caad.fkie.fraunhofer.de/details/elf.mozi} and VPNFilter\footnote{https://malpedia.caad.fkie.fraunhofer.de/details/elf.vpnfilter} include additional functionality exploiting different protocols and software vulnerabilities~\cite{Vervier2018}. The compromised devices are usually leveraged to perform different attack campaigns, including Distributed Denial of Service (DDoS) attacks, cryptocurrency mining, spamming or advertisement click fraud~\cite{Kambourakis2019}. Exposed industrial IoT (IIoT) systems are also the targets of numerous attacks that may pose additional risks due to the critical nature of these devices, including ransomware, sabotage, intellectual property theft, or be used as a pivot point to infiltrate into other systems in the IT or OT infrastructure~\cite{Sadeghi2015, McLaughlin2016}.

Several mitigation strategies have been proposed to defend against these threats. For instance, the use of specialized Operating Systems, the removal of nonessential services, reliable update mechanisms, event loggers and basic hardening operations~\cite{Kambourakis2017}. However, those mitigations do not guarantee a secure environment; misconfigurations, the discovery of new vulnerabilities and zero-days still make IoT devices prone to attacks~\cite{Meneghello2019}. As an additional security layer, Intrusion Detection Systems (IDS) and Intrusion Prevention Systems (IPS) are commonly deployed to protect the network. However, traditional signature-based IDS and other methods, such as blocklists, struggle to keep up with new IoT threats due to the usage of obfuscation techniques, packed binaries, and string modifications by malware distributors~\cite{Antonakakis2017}, rapid changes in botnet control infrastructure~\cite{Vervier2018} and long delays between the malware analysis and the publication of the corresponding rules~\cite{costin2018iot}.

Machine Learning (ML) and Deep Learning (DL) methods have shown promising results in developing IDS, exhibiting more flexibility and generalization than traditional signature-based detection methods~\cite{Ferrag2020}. However, from the point of view of ML model training infrastructure, cloud-based centralized architectures exhibit many problems in IoT settings due to the massive scale and heterogeneity of these deployments. Problems such as high bandwidth consumption, network resource congestion and load balancing arise, leading to packet loss, transmission delays, high latency and traffic peaks~\cite{Yu2018} that can adversely affect the training process or even make cloud training infeasible. In addition, data centralization can raise privacy concerns and the need to comply with regulations such as the General Data Protection Regulation (GDPR)~\cite{GDPR}. As an alternative, proposals to shift the computation toward the ``edge'' of the network are being made~\cite{Yu2018, Zhang2019}. While edge computing can alleviate some of the problems of centralized architectures, other additional issues like data islands and isolation arise, which can hinder the application of ML because it effectively reduces the volume of data available for training~\cite{Liu2021}.

A promising alternative that could address the network overhead, privacy and data isolation issues and is gaining significant attention is Federated Learning (FL). FL is a ML setting introduced in 2016 by McMahan et al.~\cite{Konecny2016} with the objective to train a single model (the global model) from data distributed at multiple remote devices (clients). The most particular characteristic about FL is that each device's local training dataset does not leave the device; instead, each client independently computes some local model update and communicates the results to a central server, which aggregates the local updates from all the clients to train the global model iteratively. Data is kept locally on each device, and only model updates are transmitted to the aggregation server, which preserves data privacy requirements. Since model updates are typically smaller than the size of the dataset, network overhead problems can also be reduced. Additionally, data isolation is minimized because multiple clients participate in training the global model.

However, there are still some difficulties to be considered for a practical FL deployment. Even though FL assumes that the data generation does not follow Independent and Identically Distributed (IID) assumptions across all the clients, in practice, highly non-IID settings can hinder global model convergence~\cite{Kairouz2019}. This can happen in highly heterogeneous settings such as large IoT networks composed of devices communicating with a diverse set of protocols.

To address the described issues, we propose a FL architecture for training anomaly-based IDS in large networks of heterogeneous IoT devices. To aggregate knowledge from all the devices, the system will leverage the FL framework to collaboratively train the anomaly detection models between multiple participants without sending each device's local data, thus reducing network overhead and tackling data isolation and privacy considerations.

In particular, to address the mentioned global model convergence problems that arise in typical FL settings with heterogeneous clients, we propose a clustered FL process that can be divided into two steps. First, before the local models are aggregated in the initial FL round, the local partially trained models from all the clients are clustered in a fully unsupervised way based on similarities between model parameters, following the hypothesis that clients with similar data distributions will converge towards models with similar parameter values. In this step, each client is assigned to a cluster center. Then, an independent FL training process is started for each identified cluster of devices. The contributions can be summarized as follows:

\begin{itemize}
    \item We propose and test a clustered FL architecture for unsupervised anomaly detection IDS model training applied to a network of heterogeneous IoT devices. We test and optimize different FL aggregation functions. The detection model is based on autoencoders trained on benign instances of IoT network traffic data to model their normal behavior. Attack traces are not used for training, only for evaluation; hence, a labeled attack dataset is unnecessary for model training.

    \item We propose an unsupervised model fingerprinting for device clustering method to address global model convergence problems in heterogeneous FL settings. The method is performed on the local model updates; thus, there is no need to send additional metadata to the FL server, incorporate external fingerprinting tools or perform manual clustering. The method is fully integrated into the FL pipeline and does not need human intervention.

    \item We evaluate the clustered FL architecture on an emulated network scenario based on the Gotham testbed~\cite{GothamTestbed}. The scenario includes 78 IoT and IIoT devices communicating with a diverse set of protocols (including MQTT, CoAP and RTSP) and different network behavior to emulate a heterogeneous environment. The IoT devices interact with 12 different cloud layer services and applications. Additionally, the scenario includes 10 attacker machines executing real IoT threats.

    \item We provide experimental results for the trained FL models. Including comparisons with a state-of-the-art approach.
\end{itemize}

The rest of this paper is structured as follows. Section~\ref{sec:relatedwork} covers the related work. Section~\ref{sec:proposedsystemmodel} discusses the proposed system model, including the FL process, the model fingerprinting for device clustering algorithm and the autoencoder anomaly detection model. Section~\ref{sec:experimentalsetup} details the IoT testbed and the data generation and collection setup. Section~\ref{sec:implementation} describes the implementation methodology of the experiments, and section~\ref{sec:results} shows the results. Finally, the paper is concluded in section~\ref{sec:conclusions}.

\section{Related Work} \label{sec:relatedwork}

In this section, we are going to describe state-of-the-art publications related to the proposals in this work. First, we will review manuscripts that apply FL techniques for IoT intrusion or anomaly detection. Then, we will describe works about clustered FL methods.

\subsection{Federated Learning for IoT Intrusion and Anomaly Detection}

Recently, several proposals have emerged that use FL techniques for IoT intrusion detection. In~\cite{Nguyen2019} Nguyen et al. present D\"Iot, an unsupervised system for network anomaly detection applied to consumer IoT devices for detecting Mirai-like worm behavior. First, an external fingerprinting tool groups all the devices based on their network behavior. Then, the FL process trains multiple global models, each one of them specific to an IoT device type group. However, one limiting factor in this approach is that a software for automatically identifying IoT device types must be available in each gateway prior to the FL process, making the model training and the device grouping not fully integrated into the same process and requiring additional time to deploy and train the system. Applied in a similar environment, Rey et al. develop in~\cite{Rey2021} a framework based on FL to detect cyberattacks against IoT devices using the N-BaIoT dataset. Additionally, they evaluate several adversarial attacks against the proposed FL framework. In~\cite{Popoola2021} Popoola et al. use the Bot-IoT and N-BaIoT datasets to train a single supervised classification global model in a FL setting and compares it with centralized and localized architectures. Another comparison between a FL intrusion detection scheme with a centralized and on-device training is shown by Rahman et al. in~\cite{Rahman2020}.

Other proposals focus on training models, or ensembles of models, that combine different input data types or views. Attota et al. propose in~\cite{Attota2021} an IDS using a multi-view ensemble of models trained with FL; three specific models are trained for each different view (network packets, unidirectional flow and bidirectional flow). Features are selected via a Grey Wolves optimization process. A random forest classifier is used to combine the prediction of the three models. Similarly, Quin et al.~\cite{Qin2021} introduce a greedy feature selection algorithm to obtain appropriate feature sets according to a single attack type that each device wants to detect. They suggest training multiple global models by grouping the devices based on the feature set selected in each client and initiating an independent FL process for each group. However, in practice, this grouping method requires prior knowledge of attacks that may not be available in a realistic environment and leaks feature set information to the aggregation server. Additionally, devices can be under multiple types of attacks at different time intervals, which will not be detected based on this method. Zhao et al.~\cite{Zhao2019} train a single multi-task model in a FL setting to perform network anomaly detection, traffic classification and Tor traffic identification simultaneously using multiple input datasets.

Alternative architectures like hierarchical FL, are also being explored for IoT intrusion detection. Wang et al.~\cite{Wang2021} describe an FL architecture based on four levels and assumes some of them are untrusted. Saadat et al.~\cite{Saadat2021} compare a standard FL architecture with a hierarchical one in terms of model training loss progression and testing accuracy for the training of an IDS using a supervised multilayer neural network on the NSL-KDD dataset. Wei et al. apply it to a 5G network~\cite{Wei2021}.

For more industrial approaches, in~\cite{Li2021} Li et al. present an IDS for industrial CPSs based on a FL scheme combined with a Paillier cryptosystem to increase the security of the model updates during the training. A recent example by Mothukuri et al.~\cite{Mothukuri2021} shows a FL-based IDS for IoT networks. They use a dataset composed of labeled network traffic data from industrial Modbus protocol. Kelli et al.~\cite{Kelli2021} propose an IDS for industrial DNP3 protocol specific attack detection combining FL and active learning to perform local model personalization for each client.

Alternative IoT attack detection approaches exist, such as UWPEE~\cite{Xie2023}, where they use a UAV to collect data from distributed IoT devices, and develop a detector based on wavelet packet energy entropy to detect attacks and assign trust to the devices. Outside of the network intrusion detection field, FL settings for IoT devices have also been proposed in sectors such as healthcare~\cite{Schneble2019, Chen2020,Huang2019} and predictive maintenance~\cite{Liu2021}, to name a few.

Most of the proposed approaches use supervised methods for model training, while unsupervised methods do not receive as much attention in the literature. In a real deployment, obtaining labeled network data to train the models is not viable at a practical level. Extending FL to unsupervised methods is still an open challenge~\cite{Kairouz2019}. Additionally, only a few papers consider the heterogeneity of IoT devices. In the cases where the heterogeneity is considered, they require a manual segmentation of the IoT devices~\cite{Schneble2019}, hardcoded device properties such as the 6-tuple in~\cite{Wei2021}, prior knowledge of attack types that target the IoT devices~\cite{Qin2021} or the help of external tools that are not fully integrated into the FL training pipeline~\cite{Nguyen2019}. Moreover, most datasets for intrusion detection were not designed to be applicable to large distributed IoT environments; therefore, many researchers resort to artificially partitioning the dataset to simulate distributed environments in which to apply FL, which is not indicative of a realistic heterogeneous IoT environment. Furthermore, most articles limit themselves to the order of 10~participating clients or less in the FL process, which does not reflect typical IoT environments, making it difficult to draw conclusions on the applicability of FL for IoT anomaly detection.

\subsection{Clustered Federated Learning}

Even though FL is based on the assumption that data is non-IID, in practice, it can show convergence problems when learning a single global model in settings with many heterogeneous clients~\cite{Kairouz2019}. Several strategies have been proposed to address these challenges and increase the personalization of FL models. One of the strategies is based on client or data clustering, which is particularly suitable for environments with inherent partitioning among FL clients, as can occur in IoT settings~\cite{Tan2022}.

Sattler et al. propose Clustered FL~\cite{Sattler2020}, which groups the client population into clusters based on the cosine similarity of clients' gradient updates. The clustering is performed as a post-processing step after the FL has converged. Ghosh et al.~\cite{Ghosh2019} present an outlier robust clustering algorithm based on K-means that also considers an adversarial setting. In~\cite{Briggs2020}, Briggs et al. introduce a clustering step to group clients according to the similarity of their local updates using hierarchical clustering methods. Then, FL is performed on each group independently.

Ghosh et al. develop IFCA~\cite{Ghosh2020}, which iteratively solves the estimation of cluster identities and model training. When the cluster structure is ambiguous, they leverage weight sharing techniques from multi-task learning. Contrary to our approach, it does not require a centralized clustering algorithm. However, since clients need to identify their own cluster membership, each client receives all k models, increasing transmission cost and client computation load. Additionally, the value of k must be known at the start of the FL process.

A Community-based FL algorithm for processing medical records is presented by Huang et al.~\cite{Huang2019}, which includes a clustering step to group the distributed data (not the clients) into several communities and a FL training step on each community. Their method requires two FL processes. The first one consists of training an autoencoder model for 1 FL round. The trained encoder is applied to each local data sample, and the averages are sent to the server to train a K-means model. In the second step, K different neural networks are trained in multiple rounds of FL. Each client receives and transmits all K models at each round. Locally, the Autoencoder and K-means are used to segment the data into K fractions, one for each global model, significantly increasing the client's workload and data transmission.

Duan et al. present FedGroup~\cite{Duan2021}, a framework that groups the clients using a proposed metric based on the cosine similarity between the optimization direction. The number of groups needs to be known a priori, and the selection of this parameter is not thoroughly discussed. Before the client clustering, a subset of the clients need to perform full pretraining of the model. After the groups are identified, the FL training process begins. In~\cite{Duan2022}, they propose an updated version that considers client distribution changes; when the shift is significant, they treat them as newcomer devices.

Xie et al.~\cite{Xie2021} propose a multi-center aggregation algorithm to learn multiple global models in a supervised learning scenario. This is performed by solving a joint optimization problem that minimizes the supervised loss function and the distance to the nearest global model of each cluster. The number of global models K needs to be known a priori, and since K is embedded into the optimization problem, selecting the optimal value of K requires repeating several FL processes fully until convergence, difficulting its application in practical settings.

Li et al.~\cite{Li2022} exploit the natural geographical clustering of factories to group the clients and propose a method that considers the divergence in class label distribution between the clients' data to minimize heterogeneity. However, the number of clusters needs to be selected prior to the training, and it requires data class labeling information, which is unsuitable for unsupervised approaches. Similarly, Hiessl et al.~\cite{Hiessl2022} group clients with similar data distributions using two approaches. The first one requires labeled data. The second one sends clients' training data statistics to the server, increasing communication costs and partially disclosing information.

Guo et al.~\cite{Guo2022} mitigate the data imbalance by presenting a data adjustment method that finds the samples corresponding to the minority class label and oversamples them. They require the FL central server to have training data to infer the data distribution of the clients and retrain the global model on the oversampled data. When the data is insufficient, the server will dynamically group clients with an adequate data class balance and use them to refine the global model at each FL round.

Other lines of work relax the hard clustering assumptions, where a client is associated with a single cluster, to a soft clustering model that allows combining data from different distributions with varying mixture ratios, as in Ruan et al.~\cite{Ruan2022}.

Our work differs from the previously referenced proposals on clustered FL in several ways. The approaches that group model parameters using centralized clustering algorithms~\cite{Sattler2020,Ghosh2019,Briggs2020} lead to high computation costs and may not be practical for setting with large models and a large number of clients. In contrast, we include parameter dimensionality reduction methods to mitigate that issue. Other proposals require each client to process K models~\cite{Ghosh2020,Huang2019}, increasing local computation and bandwidth load. In~\cite{Ghosh2020,Duan2021,Duan2022,Xie2021,Li2022}, the number of clusters needs to be known a priori, and selecting an optimal value for it requires completing the full clustered FL training, which is costly and complicates the hyperparameter selection step in practical settings. More importantly, all the approaches assume a supervised learning setting, and some require the presence of labels to perform the clustering or personalization process~\cite{Li2022,Guo2022}. In contrast, our proposal and experimentation methodology focus on unsupervised settings. Finally, none of those methods were applied to the IoT security field. This paper presents a practical use of clustered FL methods to address IoT network attacks using clustering methods fully integrated into the FL training pipeline and unsupervised anomaly-based network intrusion detection models.

\section{Proposed System Model} \label{sec:proposedsystemmodel}

This section first shows a high-level overview of the proposed system architecture and the targeted deployment setting. Then, we present the proposed clustered FL architecture, describing our contributions on top of the standard FL process to include the integrated model fingerprinting for the device clustering step. We finally have a brief review of autoencoder neural networks for anomaly detection.

\subsection{Deployment Setting and Architecture}

\begin{figure}
	\centering
	\includegraphics{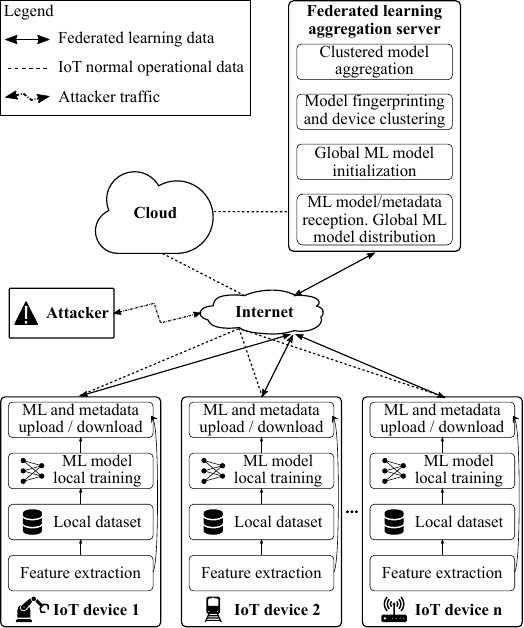}
	\caption{Proposed system architecture. Each IoT device (FL clients) holds a copy of the ML model for local training and inference. The FL training process is mediated by the aggregation server. The FL aggregation server can also be part of the IoT cloud, but here it is shown separately for clarity.}
	\label{fig:architecture}
\end{figure}

The proposed architecture to train the IDS is depicted in \figurename~\ref{fig:architecture}. It comprises many clients and a central aggregation server and also shows potential attackers.

\subsubsection{Clients}
The proposed system is devised to operate in a large network of heterogeneous IoT devices such as gateways, CPS and industrial machines that communicate using different protocols. Those devices can be located in different network segments or geographically distributed, which may influence their behavior. The devices are constantly sending/receiving data to/from the cloud layer.

Each device is considered a client in the FL process. They are responsible for capturing relevant data, local ML model training, and model inference for anomaly detection after the training is completed. No training data is transmitted to the aggregation server, only model parameters and minimal metadata relevant to the FL process. Devices are expected to perform lightweight ML tasks, but for low-powered IoT devices such as sensors and actuators, the system is expected to be deployed at the hub or gateway level. In this work, we prioritize the use of lightweight ML models for anomaly detection to limit the computational overhead during model training or inference.

\subsubsection{Aggregation server}
It coordinates all the FL training process by initializing and distributing the model and training hyperparameters to the clients, receiving model updates from the clients, performing the model fingerprinting and device clustering, running the per cluster aggregation of the received models and sending the corresponding aggregated global model to each client. These steps are explained in the next subsection.

\subsubsection{Attackers}
In this paper, we consider two primary threat models. The first one considers external actors that remotely scan the IoT devices in the network, find vulnerable devices to exploit and remotely compromise them. The second assumes a local adversary compromising one or many IoT devices within the protected network and leveraging them to launch attacks against other devices in the same network or target external victims. In section~\ref{sec:experimentalsetup}, we will detail the different threat actors, attacks performed and malicious behavior.

\subsection{Clustered Federated Learning Process for Heterogeneous Devices}

For the FL deployment, we consider a typical cross-device setting~\cite{Kairouz2019} with a large number of devices. However, due to availability guarantees required by many IoT devices, especially in industrial settings, we expect most devices to participate in the FL training process. This allows the server to maintain a persistent state for each client to perform the clustering step. In this work, we assume that no IoT device is infected prior to the FL model training and that none of them behave in an adversarial manner. Model poisoning attacks against FL are outside the scope of this paper, even if already considered in some academic works~\cite{Bagdasaryan2020}.

Our clustered FL builds upon the generalized FL setting proposed by Reddi et al.~\cite{Reddi2021}. This generalized FL setting, described in Algorithm~\ref{alg:fl}, improves over standard FL optimization methods such as the popular Federated Averaging (FedAvg)~\cite{McMahan2017} by including adaptive optimization methods for the local model training at each client and also at the server level during the model aggregation process.

\begin{figure}[!t]
\removelatexerror
\begin{algorithm}[H]
\SetAlgoLined
\SetKw{KwIn}{in}
\SetKw{KwInparallel}{in parallel}
\SetKwInput{KwInput}{Input}
\DontPrintSemicolon

 \SetKwFunction{FLocalTrain}{LocalTrain}
 
 \SetKwProg{Fn}{Function}{:}{}

  \Fn{\FLocalTrain{$\mathbf{w}$, epochs}}{
    \For{local epoch e $\leftarrow$ 1 \KwTo epochs}{
        \For{batch b \KwIn local training data}{
            $\mathbf{g} \leftarrow$ compute gradient\;
            $\mathbf{w} \leftarrow$ \textsc{ClientOpt}($\mathbf{w}$, $\mathbf{g}$, $\eta$, e)\;
        }
    }
    \KwRet $\mathbf{w}$, number of local training samples\;
 }
 \;
 
 \KwInput{A set of clients $\cal C$, initialized model $\mathbf{W_0}$}
 \KwResult{Trained global model $\mathbf{W_G}$}
 
 E $\leftarrow$ number of local epochs\;
 R $\leftarrow$ total federated learning rounds\;
 $\mathbf{W_G} \leftarrow \mathbf{W_0}$\;
 \For{federated learning round t = 1, 2, \ldots, R}{
    \ForEach{client $c \in \cal C$ \KwInparallel}{
        receive $\mathbf{W_G}$ from the server\;
        $\mathbf{W_c}, n_c \leftarrow$ \FLocalTrain{$\mathbf{W_G}$, E}\;
        local model delta $\Delta_c \leftarrow \mathbf{W_c} - \mathbf{W_G}$\;
        send $(\Delta_c, n_c)$ to the server\;
    }
    
    $n \leftarrow \sum_{i \in \cal C} n_i$\;
    pseudogradient $\mathbf{g_G} \leftarrow -\sum_{i \in \cal C} \frac{n_i}{n} \Delta_i$\;
    $\mathbf{W_G} \leftarrow $ \textsc{ServerOpt}($\mathbf{W_G}$, $\mathbf{g_G}$, $\eta_s$, $t$)\;
 }

 \caption{Generalized federated learning process. The \textsc{ClientOpt}, \textsc{ServerOpt}, their respective learning rates ($\eta$, $\eta_s$) and the pseudogradient concepts are explained in detail by Reddi et al.~\cite{Reddi2021}}
 \label{alg:fl}
\end{algorithm}
\end{figure}

\begin{figure}[!t]
\removelatexerror
\begin{algorithm}[H]
\SetAlgoLined
\SetKw{KwIn}{in}
\SetKw{KwInparallel}{in parallel}
\SetKwInput{KwInput}{Input}
\DontPrintSemicolon

 \SetKwFunction{FModelFingerprinting}{ModelFingerprinting}
 \SetKwFunction{FLocalTrain}{LocalTrain}
 \SetKwProg{Fn}{Function}{:}{}
 
 \Fn{\FModelFingerprinting{$weight\_list$}}{
  $\cal W \leftarrow $ empty list \;
  
  \For{$\mathbf{w}$ \KwIn $weight\_list$}{
   append flattened $\mathbf{w}$ to $\cal W$\;
  }
  
  $\cal W \leftarrow $ apply PCA dimensionality reduction to $\cal W$\;
  
  $\cal S \leftarrow$ empty list \;
  $\cal L \leftarrow$ empty list \;
  \For{n $\leftarrow$ 2 \KwTo max number of clusters}{
   K-means clustering of $\cal W$ into n clusters\;
   append clustering labels to $\cal L$\;
   append clustering quality score to $\cal S$
  }
  
  $K \leftarrow$ number of clusters with optimal score in $\cal S$\;
  
  \KwRet labels from $\cal L$ corresponding to $n=K$, $K$\;
 }
 \;

 \KwInput{A set of clients $\cal C$}
 \KwResult{A set of global models}
 initialize model $\mathbf{W_0}$ on server\;
 $\epsilon \leftarrow$ number of local epochs for clustering\;
 
 \ForEach{client c $\in \cal C$ \KwInparallel}{
    receive $\mathbf{W_0}$ from the server\;
    $\mathbf{W_c}, n_c \leftarrow$ \FLocalTrain{$\mathbf{W_0}$, $\epsilon$}\;
    send $\mathbf{W_c}$ to the server
 }
 $\cal W \leftarrow$ list of all the received $\mathbf{W_{c \in \cal C}}$ \;
 
 $\cal L$, $K$ $\leftarrow$ \FModelFingerprinting{$\cal W$}\;
 
 \ForEach{label $k \in \{1,...,K\}$ \KwInparallel}{
    ${\cal C}_k$ $\leftarrow$ subset of clients $\in \cal C$ with labels ${\cal L} = k$\;
    $\mathbf{W_{G}^{C=k}} \leftarrow$ average of $\cal W$ with labels ${\cal L} = k$\;
    $\mathbf{W_{G}^{C=k}} \leftarrow $\texttt{FederatedLearning}$({\cal C}_k, \mathbf{W_{G}^{C=k}})$\;
 }

 \caption{Proposed clustered federated learning for heterogeneous clients. The \texttt{LocalTrain} and \texttt{FederatedLearning} functions are described in Algorithm~\ref{alg:fl}.}
 \label{alg:heterogeneousfl}
\end{algorithm}
\end{figure}

The proposed clustered FL is described in Algorithm~\ref{alg:heterogeneousfl}. First, the aggregation server initializes the model weights~$\mathbf{W_0}$ and selects the training hyperparameters. Then, the server sends those values to all the participating clients. In the next step, each client partially trains $\mathbf{W_0}$ using only its local data for $\epsilon$~epochs. The local training is performed using the \textsc{ClientOpt}~\cite{Reddi2021} gradient-based optimizer to minimize the local training loss. \textsc{ClientOpt} is an abstraction for optimizers such as SGD, Adam or RMSprop. After the local training, each client sends the partially trained model to the aggregation server. The aggregation server collects all the local models and uses them to group the clients into $K$~clusters based on similarities between the trained model parameters (weights and biases). The grouping process is discussed in more detail in the next subsection.

For each identified cluster~$k$, an independent FL process is executed in parallel (Algorithm~\ref{alg:fl}). We perform multiple FL rounds (R rounds) until the global model for each cluster converges, resulting in a set of $K$~global models. At each round, the clients transmit the difference between the weights from the received global model at the start of the round and the locally updated model weights. The server uses these weight deltas to compute what the authors in~\cite{Reddi2021} call as \textit{pseudo-gradient}, i.e., the negative of the averaged model deltas. The \textit{pseudo-gradient}, along with the server learning rate $\eta_s$ is used for the model aggregation process, which is generalized in the \textsc{ServerOpt} function as shown in Algorithm~\ref{alg:fl}. The \textsc{ClientOpt} and \textsc{ServerOpt} abstraction allows incorporating momentum or other adaptive optimization methods to both client-side training and server-side model aggregation compared to the FedAvg algorithm~\cite{Reddi2021}. The popular FedAvg aggregation method can be considered a special case where \textsc{ServerOpt} is set to SGD with server learning rate~$\eta_s = 1.0$.

\subsection{Model Fingerprinting for Device Clustering}

In a network of heterogeneous devices, the underlying data distribution might not be IID. In a FL setting, a single global model complex enough could be able to fit the data properly; however, training a complex model in IoT devices might not be possible due to hardware constraints. Consequently, we will group the devices with similar behavior to create a set of global models specifically tailored to each group of devices. With this method, each IoT device is assigned a group label in an unsupervised manner that is going to be used during the FL process.

The main advantages of using the locally trained model updates as inputs for the clustering method are that \textit{i)} there is no need to integrate any external device fingerprinting algorithms or manual methods, \textit{ii)} does not require waiting for a certain amount of time to identify the devices before starting the model training process and \textit{iii)} everything is completely integrated into the FL training pipeline.

As detailed in Algorithm~\ref{alg:heterogeneousfl}, the first step for the model fingerprinting consists of partially training each local model for $\epsilon$~epochs, and sending the partially trained model to the aggregation server. Then, the server flattens the parameters (weights and biases) of each model and performs Principal Component Analysis (PCA) to reduce the dimensionality of the parameters. The reduced dimensionality helps speed up the computation of the clustering step and can limit the problems of clustering high dimensionality data in models with a considerable number of parameters. We use the K-means algorithm with the k-means++ initialization scheme~\cite{Arthur2006} to cluster the reduced dimensionality data. The hypothesis is that clients with similar data distributions will converge to models with similar parameter (weights and biases) values, provided that all clients start from the same initial random model~$\mathbf{W_0}$.

Due to the unsupervised nature of our proposal, we will use internal clustering validation metrics to select an optimal value for the number of clusters~$K$. Internal validation metrics do not rely on any external data and are mainly based on measures such as the compactness of samples within the same cluster and separation between different clusters~\cite{Liu2010}. Specifically, we will evaluate the following internal validation metrics: Silhouette~\cite{Rousseeuw1987}, Davies-Bouldin~\cite{Davies1979} and S\_Dbw~\cite{Halkidi2001} scores to select the value of~$K$.

\subsection{Anomaly Detection Model}

In this paper, we are going to employ autoencoder neural networks as the anomaly detection models, which have already been used in similar domains for network-based attack detection~\cite{Mirsky2018,Meidan2018}. We are going to prioritize lightweight autoencoders (small number of parameters), which makes them especially suitable for our deployment scenario because it not only requires less computational load for model training or inference in constrained devices, but also reduces the network traffic volume between the devices and the aggregation server during the FL rounds due to less number of parameters compared to more complex models.

Autoencoders are unsupervised neural networks that attempt to replicate the input data on their output layer under some constraints to avoid learning the identity function. The autoencoder is composed of two networks, the encoder and the decoder. The encoder takes the input features~$\mathbf{x} \in \mathbb{R}^n$ and transforms them into a hidden encoded space ~$\mathbf{h} \in \mathbb{R}^e$, where~$e < n$ to impose a constraint to avoid learning the identity function. Then, the decoder transforms~$\mathbf{h}$ into~$\mathbf{x^\prime} \in \mathbb{R}^n$. The objective of the autoencoder is to minimize the mean squared error (MSE, reconstruction error) between~$\mathbf{x}$ and~$\mathbf{x^\prime}$ as in equation~\eqref{eq:mse}. The autoencoder is trained using the loss function shown in equation~\eqref{eq:loss_function}, which in addition to the MSE, it includes the~$L_2$ regularization term.

\begin{equation}
	MSE = \frac{1}{n} \sum_{i=1}^{n} (x_i - x_i^\prime)^2.
	\label{eq:mse}
\end{equation}

\begin{equation}
	\mathcal{L} = MSE + \lambda \sum_{i} w_{i}^{2}
	\label{eq:loss_function}
\end{equation}

We train the autoencoder using samples of normal (legitimate or benign) IoT network traffic which does not contain attacks; this way, the model learns a representation of the normal behavior of these devices. Once the autoencoder is trained (using the proposed FL approach), it is evaluated on network traces containing legitimate and attack samples. The reconstruction error between the input and output layers is used as a measure of the anomaly level in the new incoming data. New network samples that came from a similar distribution as the training data will have a small reconstruction error; however, we expect that attack samples diverge from the trained data distribution, and thus, the reconstruction error will be higher.

\section{IoT Testbed and Experimental Setup} \label{sec:experimentalsetup}

In this section, we present the experimental setup. We begin with a description of the selected IoT testbed used to extract the dataset and implement the experiments, including a description of all the considered IoT device types and attackers. We then continue to detail the training and validation network traffic datasets generated with the testbed. For the validation datasets, we include a list of the malicious activities performed according to our threat model.

\subsection{IoT Testbed}

The experimental setup is based on the Gotham Testbed~\cite{GothamTestbed}, a testbed to perform large-scale IoT security experiments in a realistic and reproducible way. Gotham is built on top of the GNS3~\cite{Grossmann} network emulator, and it includes a repository of Docker images and QEMU-based virtual machines to emulate various IoT/IIoT devices, malware samples, servers and networking equipment such as switches and routers. To generate real network traffic traces, the emulated devices run real production libraries, network switching software (Open vSwitch) and routing operating systems (VyOS), as well as real malware samples and red-teaming tools, which are going to be briefly described in this section.

To create the IoT scenario, we use the default case study scenario presented at~\cite{GothamTestbed} composed of many heterogeneous nodes. The scenario comprises three main networks connected by 10 routers and 30 switches: the city network, the cloud network and the threat network. The full details are explained in~\cite{GothamTestbed}, but here we will present a summary of the different emulated devices which are relevant for the discussion in the following sections.

\subsubsection{City network devices}

The city network contains the emulated IoT/IIoT devices. The devices communicate with the cloud network using various protocols, including MQTT, CoAP (two protocols that are specifically designed and well-suited for machines with constrained resources and the IoT paradigm~\cite{Minerva2015}) and RTSP. Additionally, to increase the protocol heterogeneity, the devices generate background traffic such as DNS, NTP and ICMP. In total, there are 12 device templates to simulate a heterogeneous environment, as shown in Table~\ref{tab:iot_device_templates}.

\begin{table}[t]
	\small
	\centering
	
	\caption{Included IoT/IIoT device templates in the testbed scenario.}
	\label{tab:iot_device_templates}
	
	\begin{tabularx}{1.0\linewidth}{p{0.34\linewidth}|p{0.13\linewidth}|X}
		\toprule
		Template name          & Instances & Main protocol \\
		\midrule
		Air quality            & 1         & MQTT (plain)      \\
		Building monitor       & 5         & MQTT (plain)      \\
		City power             & 1         & CoAP (plain)      \\
		Combined cycle         & 10        & CoAP (plain)      \\
		Combined cycle tls     & 5         & CoAP (DTLS)      \\
		Cooler motor           & 15        & MQTT (plain and TLS)      \\
		Domotic monitor        & 5         & MQTT (plain)      \\
		Hydraulic system       & 15        & MQTT (plan and TLS)      \\
		IP camera street       & 2         & RTSP      \\
		IP camera museum       & 2         & RTSP      \\
		IP camera consumer     & 2         & RTSP      \\
		Predictive maintenance & 15        & MQTT (plain and TLS)      \\
		\bottomrule
	\end{tabularx}
	
\end{table}

Each device template presented in Table~\ref{tab:iot_device_templates} has a distinct behavior. The telemetry payload size (from 10~bytes per payload to more than 7500~bytes per payload), format (JSON, XML, Base64) and periodicity of the communications vary between the templates. The telemetry data is transmitted in plain text for some nodes and over an encrypted channel using TLS or DTLS for other nodes. Some brokers at the cloud network accept unauthenticated clients, while others require clients to be authenticated with a username and a password before sending the data. For the MQTT-based nodes, some templates only publish to a single topic, while others publish to multiple topics. The CoAP-based nodes also serve different numbers of resources, nine for the City power and five resources for the Combined cycle nodes. Regarding data transmission behavior, some devices open a single connection with the cloud at the beginning of the transmission and keep it alive by periodically sending telemetry data and keep-alive messages. Other nodes open a new connection to the cloud, send the data and then close the connection each time they need to send telemetry.

The Gotham testbed scenario includes a total of 78~instances of those device templates, as shown in Table~\ref{tab:iot_device_templates}. To increase the heterogeneity, each instance has small random deviations and jitter following a normal distribution in the periodicity of the communications.

The distributed dataset generated by the testbed is highly non-IID due to all the different client behaviors implemented in it. It primarily includes high feature distribution skew and data quantity skew (all clients do not generate the same amount of data). Additionally, due to the data being network traces, samples can be non-independent. The described behaviors are some of the common ways in which data is non-IID, as described by Kairouz et al.~\cite{Kairouz2019}.

\subsubsection{Cloud network devices}

The cloud network includes the necessary cloud services to enable communication with the devices in the city and threat networks. The services include many MQTT brokers, CoAP clients and IP camera streaming servers. Additionally, the cloud network includes nodes to provide DNS and NTP services.

\subsubsection{Threat network devices}

To launch realistic attacks, the testbed scenario includes three threat actors.

\paragraph{Maroni Crime Family}
It includes the real Mirai~\cite{Antonakakis2017} malware. All the nodes in this threat actor were created based on the published Mirai source code~\cite{Miraisrc}, adapted and compiled to run on the testbed. The nodes include the (i) Mirai bot, (ii) Mirai Command and Control (C\&C) server, (iii) Mirai scan listener, (iv) Mirai loader and the (v) Mirai download server. All the nodes in this threat actor allow the execution of the whole Mirai malware lifecycle. We use this threat actor to perform the following attack activities:

\begin{itemize}
	\item \textbf{(A1) Mirai C\&C communication}: Includes the periodic communication between the Mirai bots and the Mirai C\&C server.
	
	\item \textbf{(A2) Mirai network scanning}: Each bot infected with Mirai scans the network in a pseudorandom order sending TCP SYN packets to the Telnet 23 and 2323 ports.
	
	\item \textbf{(A3) Mirai brute forcing}: If the Mirai scanner detects an open telnet port, it tries to brute force the credentials using a list of common IoT usernames and passwords.
	
	\item \textbf{(A4) Mirai reporting}: After a successful brute forcing, the Mirai bot reports the victim's IP address, port, username and password to the Mirai scan listener.
	
	\item \textbf{(A5) Mirai ingress tool transfer}: Includes the infection phase of Mirai. The Mirai loader connects to vulnerable nodes listed in the Mirai scan listener server and proceeds to download and execute the malware.
	
	\item \textbf{(A6) Mirai remote command execution}: The Mirai bot master connects to the Mirai C\&C and sends commands to the bots to perform subsequent DoS attacks against other targets in the network.
	
	\item \textbf{(A7) Mirai denial of service attacks}: The following list enumerates the performed DoS attacks by the Mirai bots against the targets (it does not include all attack types supported by Mirai): (i) UDP plain attack, (ii) UDP attack, (iii) Valve Source engine attack, (iv) DNS attack, (v) TCP ACK attack, (vi) TCP SYN attack, (vii) GRE IP attack, (viii) GRE Ethernet attack. All attacks were performed for a duration of 10s. All attacks targeted other IoT devices in the city network, except for the DNS attack, which targeted the DNS server at the cloud network.
\end{itemize}

\paragraph{Falcone Crime Family}
This threat actor includes the Merlin~\cite{Tuyl} cross-platform post-exploitation C\&C server, the Merlin agents and the hping3~\cite{Sanfilippo} TCP/IP packet assembler and analyzer. The Merlin server supports multiple protocols for C\&C (HTTP/1.1 clear-text, HTTP/1.1 over TLS, HTTP/2, HTTP/2 clear-text (h2c), HTTP/3) and can execute code on the victims under its control. When a victim device is compromised with the Merlin agent, it starts communicating with the Merlin C\&C and becomes part of its botnet. We use this threat actor to perform the following attacks:

\begin{itemize}
	\item \textbf{(A8) Merlin C\&C communication}: Periodic communication between the IoT nodes infected with the Merlin agent and the Merlin C\&C server.
	
	\item \textbf{(A9) Merlin ingress tool transfer}: The Merlin C\&C server transfers the hping3 binary into each of the compromised victims through the C\&C channel.
	
	\item \textbf{(A10) Merlin remote command execution}: The Merlin bot master connects to the Merlin C\&C and sends commands to the bots to perform subsequent DoS attacks against other targets in the network.
	
	\item \textbf{(A11) Merlin denial of service attacks}: The DoS attacks are implemented using hping3: (i) ICMP echo-request, (ii) UDP, (iii) TCP SYN and (iv) TCP ACK flood attacks. Each attack generates approximately 5000~packets at a 1~ms/packet rate. The UDP flood payload consists of 512~bytes of random data, with TTL set to 64 and TOS to 0, which corresponds to the default values for the UDP attack in Mirai.
\end{itemize}

\paragraph{Calabrese Crime Family}
It includes many nodes performing network-wide scanning operations and launching attacks against the devices. It is comprised of nodes that include the Nmap and Masscan scanners and the AMP-Research tool for implementing amplification attacks against the CoAP servers. We use this threat actor to perform the following attacks:

\begin{itemize}
	\item \textbf{(A12) Network-wide scans}: Masscan is used to scan the city network for the TCP ports 80,8000-8100,5683 at three different packet rates: (i) 100, (ii) 1000 and (iii) 10000~packets/s. Nmap is used to scan some IoT nodes to check for open UDP ports using three strategies: (iv) the 5683 port, (v) 600 random ports and (vi) the 1000 most used UDP ports.
	
	\item \textbf{(A13) CoAP amplification attack}: The attacker leverages a CoAP device in the city network to launch an amplification attack against a victim for a duration of 10s.
\end{itemize}

\subsection{Data Generation and Collection Method} \label{sec:data-generation-collection}

The Gotham testbed allows the capturing of network traffic traces for dataset generation. The data captured under normal traffic conditions (without attacks) will be used to train the clustered FL anomaly detection models. Then, the validation data is captured, consisting of two sets: validation-normal and validation-attack. The validation-attack is further divided into different datasets depending on the attack scenario.

All this data is captured in a federated way. Each device holds its own part of the data (captured on its network interface), as shown in \figurename~\ref{fig:architecture}. This data is never aggregated into a single dataset.

\subsubsection{Normal traffic data}

This data is composed of the normal behavior of the city network IoT/IIoT devices periodically communicating the telemetry and background data with the cloud. Network packet traces are collected for each device and saved in pcap format while the scenario runs without any attack. This dataset will be used for feature preprocessing, hyperparameter selection and the clustered FL training.

The normal traffic data has been captured in a period including the first two hours (the first hour for the IP camera related devices, due to the high data volume they create), generating a total of 3.3~GB of raw packet data for all the 78~devices in the network. 

\subsubsection{Validation-normal traffic data}

The validation-normal dataset consists of traces including only the normal behavior of the city network devices captured with the same methodology from the previously described normal dataset. However, it is extracted later so that it does not include the same events. It includes captures over a two-hour period (one hour for the IP cameras) starting after the end of the normal traffic capture. This data is not used during training; it will only be used for the anomaly threshold selection.

\subsubsection{Validation-attack traffic data}

While the city network devices are performing their normal activities, the attacker nodes become active and start launching the previously mentioned attacks against the city network IoT/IIoT devices. The validation-attack traffic data is captured during this period and consists of both normal and attacking traces.

We configure the testbed's three threat actors to create five attacking scenarios. For each scenario, we extract network packet captures from the city network devices.

\paragraph{validation-attack-mirai-scan-load}

Some city network devices are first configured to make them vulnerable to Mirai, as detailed in~\cite{GothamTestbed}. The testbed's Mirai bot node is activated (\textbf{A1}) and starts scanning all city network devices (\textbf{A2}). When vulnerable devices are identified, the Mirai bot performs the (\textbf{A3}) and (\textbf{A4}) activities. After a vulnerable device is reported, the (\textbf{A5}) activity is performed to integrate the device into the botnet. After becoming part of the botnet, the device repeats the described Mirai lifecycle.

\paragraph{validation-attack-mirai-cnc-dos}

This data also includes Mirai malware activity, but in this case, we recompile the Mirai bot binary to disable the scanning and brute-forcing modules. This modification is done to make Mirai stealthier. The modified Mirai bot is manually installed in some city network devices. After executing the bot, the C\&C communication activity starts (\textbf{A1}). By connecting to the Mirai C\&C server, we command each bot (\textbf{A6}) to launch multiple DoS attacks (\textbf{A7}) against random targets in the testbed.

\paragraph{validation-attack-merlin-cnc-dos}

The Merlin agent is installed in some city network devices. After executing the agents, they connect to the Merlin server (\textbf{A8}). For each bot, the Merlin server performs (\textbf{A9}) and (\textbf{A10}). Finally, each bot is instructed to launch DoS attacks (\textbf{A11}) against random targets in the testbed.

\paragraph{validation-attack-masscan}

Network traffic data is captured from the city network devices while they are being scanned by the Masscan node (\textbf{A12}).

\paragraph{validation-attack-scan-amplification}

This data is captured on the CoAP-based city network devices. First, Nmap is used to scan the network (\textbf{A12}) to search for CoAP devices; then, those devices are leveraged to perform (\textbf{A13}) attacks against random targets in the testbed.

\subsection{Machine Learning and Federated Learning Setup}

We used the PyTorch~\cite{Paszke2019} Python library to implement the ML models and training procedures. The FL model aggregation (the \textsc{ServerOpt} server-side optimization) is also implemented using the PyTorch library directly. For the client's local training process, we used GNU Parallel~\cite{Tange2011} to coordinate and execute all the jobs in parallel. We implemented the clustering algorithms, validation metrics, dimensionality reduction, etc. with scikit-learn~\cite{scikit-learn}.

\section{Implementation} \label{sec:implementation}

\begin{figure*}
	\centering
	\includegraphics[width=1.0\linewidth]{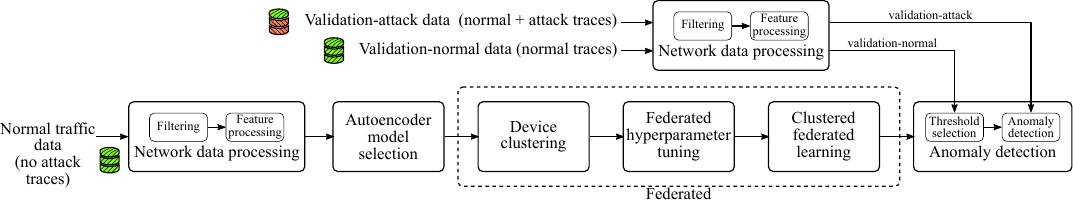}
	\caption{Implementation method pipeline.}
	\label{fig:implementation_pipeline}
\end{figure*}

This section will describe the methodology followed to perform the experimentation. A visual representation of all the steps is shown in \figurename~\ref{fig:implementation_pipeline}. We first explain the network data processing step, which includes filtering, feature extraction and preprocessing. Then, we detail the autoencoder model selection procedure. Next, we describe the implementation for the clustered FL process, starting from the model fingerprinting for device clustering, followed by the federated hyperparameter tuning and then the FL training process for each identified cluster. Finally, we review the trained models' anomaly detection evaluation process and metrics. We additionally explain the baseline comparisons done with other state-of-the-art IDS methods.

\subsection{Network Data Processing}

After collecting the dataset, the raw pcap files are first filtered, then relevant network features are selected, and finally, those features are preprocessed to make them suitable as input to the ML models. Note that the dataset is federated and not centralized; each device holds its fraction of data.

The first step consists of filtering the raw pcap files to drop all IPv6 and ARP packets. The filtered packets are passed to the feature extraction process. For each network packet in the filtered pcap file, a set of 11~features are extracted as listed in Table~\ref{tab:packet_features_description}. The source and destination IP addresses were discarded to prevent the model from learning the machines themselves instead of the attacking nature. The main reason for using those features is that attacking patterns from IoT malware such as Mirai includes options to craft packets with tweaked values for the payload size, IP header fields, and TCP flags, among others~\cite{Miraisrc}. Additionally, the attack packet payload usually includes randomized or some fixed values that can lead to high or low entropy. By selecting those features and training on the normal traffic dataset, the model learns the distribution of normal IoT communication. Deviations from it (large MSE between the input and the autoencoder reconstructed output) allows to potentially detect not only Mirai but, in general, other malware with similar network attacking behavior and C\&C communication like the attacks performed with hping3 and Merlin C\&C as described in the previous section.

Due to the different orders of magnitude of some features and the mixture of both numerical and categorical variables, a feature preprocessing step is necessary before using them as inputs for the ML models.

\begin{table}[t]
    \small
    \centering
    
    \caption{Selected packet feature names and descriptions.}
    \label{tab:packet_features_description}

    \begin{tabularx}{1.0\linewidth}{l|l}
    \toprule
        Feature name & Description \\
    \midrule
        \texttt{len} & Full packet length in bytes. \\
        \texttt{iat} & Inter arrival time from the previous packet. \\
        \texttt{h} & Entropy (base 2) of the full packet. \\
        \texttt{ip\_tos} & IP type of service. \\
        \texttt{ip\_flags} & IP flags (MF, DF, R bits). \\
        \texttt{ip\_ttl} & IP time to live. \\
        \texttt{ip\_proto} & IP protocol (TCP, UDP, ICMP). \\
        \texttt{src\_port} & Source port number. \\
        \texttt{dst\_port} & Destination port number. \\
        \texttt{tcp\_flags} & TCP flags (F, S, R, P, A, U, E, C, N). \\
        \texttt{tcp\_win} & TCP window size. \\
    \bottomrule
    \end{tabularx}

\end{table}

The numerical features are normalized by the maximum value of each field defined in the TCP/IP stack. \texttt{len} is divided by 1514 (the Ethernet maximum transmission unit plus the header), both \texttt{ip\_tos} and \texttt{ip\_ttl} are divided by 255 and \texttt{tcp\_win} is divided by 65535. \texttt{h} is divided by 8 and \texttt{iat} is transformed with the natural logarithm of \texttt{iat} plus one. The categorical variables \texttt{ip\_flags}, \texttt{ip\_proto} and \texttt{tcp\_flags} are directly one-hot encoded.

\subsubsection{Source and destination port feature processing}

The \texttt{src\_port} and \texttt{dst\_port} feature processing requires special consideration. The application port numbers are numerical features that can take $2^{16}$ different integer values; however, treating the ports just as a numerical feature does not maintain the semantics of the services that use those port numbers. In other words, port numbers that are numerically close to each other does not mean that the programs that communicate with those ports are used to perform similar tasks.

In this work, we are going to discretize the source and destination port numbers into a smaller number of bins using two different strategies: three-range discretization and hierarchical discretization. After the discretization, the bin numbers are one-hot encoded. In section~\ref{sec:results}, we will evaluate the differences between the two strategies and select the most appropriate one for this use case.

\paragraph{Three-range discretization}

The source and destination port numbers are divided into the three ranges assigned by the Internet Assigned Numbers Authority (IANA)~\cite{RFC6335}: the System Ports, or Well Known Ports, from 0 to 1023; the User Ports, or Registered Ports, from 1024 to 49151; and the Dynamic Ports, or Ephemeral Ports, from 49152 to 65535. These ranges are large, and might not capture the semantics of the ports.

After all the previously mentioned transformations and using the three-range discretization for the source and destination ports, the 11~features of Table~\ref{tab:packet_features_description} are transformed into a set of 27~features.

\paragraph{Hierarchical discretization}

The source and destination port numbers are discretized based on the generalization hierarchy presented in~\cite{Zurutuza2008} and adapted to include information about the MQTT, CoAP and RTSP ports used in the testbed. The hierarchy used is summarized in Table~\ref{tab:port_hierarchichal_discretization}. When classifying a port number, bins from the top of the table have precedence over the bins from the bottom.

After all the previously mentioned transformations and using the hierarchical discretization for the source and destination ports, the 11~features of Table~\ref{tab:packet_features_description} are transformed into a set of 69~features.

\begin{table}[t]
	\small
	\centering
	
	\caption{Port number generalization hierarchy.}
	\label{tab:port_hierarchichal_discretization}
	
	\begin{tabularx}{1.0\linewidth}{p{0.27\linewidth}|X}
		\toprule
		Bin name           & Ports                                                            \\
		\midrule
		mqttPorts          & 1883, 8883                                                     \\
		coapPorts          & 5683, 5684                                                     \\
		rtspPorts          & 8554, 8322, 8000 -- 8003, 1935, 8888                 \\
		httpPorts          & 80, 280, 443, 591, 593, 777, 488, 1183, 1184, 2069, 2301, 2381, 8008, 8080  \\
		mailPorts          & 24, 25, 50, 58, 61, 109, 110, 143, 158, 174, 209, 220, 406, 512, 585, 993, 995 \\
		dnsPorts           & 42, 53, 81, 101, 105, 261                                           \\
		ftpPorts           & 20, 21, 47, 69, 115, 152, 189, 349, 574, 662, 989, 990                    \\
		shellPorts         & 22, 23, 59, 87, 89, 107, 211, 221, 222, 513, 614, 759, 992                 \\
		remoteExecPorts    & 512, 514                                                        \\
		authPorts          & 13, 56, 113, 316, 353, 370, 749, 750                                  \\
		passwordPorts      & 229, 464, 586, 774                                                \\
		newsPorts          & 114, 119, 532, 563                                                \\
		chatPorts          & 194, 258, 531, 994                                                \\
		printPorts         & 35, 92, 170, 515, 631                                              \\
		timePorts          & 13, 37, 52, 123, 519, 525                                           \\
		dbmsPorts          & 65, 66, 118, 150, 156, 217                                          \\
		dhcpPorts          & 546, 547, 647, 847                                                \\
		whoisPorts         & 43, 63                                                          \\
		netbiosPorts       & 137 -- 139                                                 \\
		kerberosPorts      & 88, 748, 750                                                     \\
		RPCPorts           & 111, 121, 369, 530, 567, 593, 602                                    \\
		snmpPorts          & 161, 162, 391                                                    \\
		privilegedPorts    & 0 -- 1023                                                    \\
		nonprivilegedPorts & 1024 -- 65535                                                \\
		\bottomrule
	\end{tabularx}
	
\end{table}

\subsection{Autoencoder Model Selection}

In FL, there is a much larger set of hyperparameters to be tuned compared to a typical centralized ML setting. Those parameters include the ML model itself (number of layers, number of nodes per layer, activation functions, etc.), client-side optimizer algorithm \textsc{ClientOpt} and learning rate~$\eta$, server-side optimizer algorithm \textsc{ServerOpt} and learning rate~$\eta_s$, number of local training epochs~$E$, number of FL rounds~$R$ and number of clients sampled per FL round~$M$. Due to the infeasibility to explore all the combinations simultaneously, we are going to simplify the search tuning those hyperparameters step by step using different subsets of combinations. Additionally, considering the unsupervised nature of the problem (or rather semi-supervised, given that it is trained on normal data without attacks), the selection is going to be based on those values that minimize the MSE loss in fewer rounds/epochs.

First, we start defining the general architecture of the autoencoder. We select a small subset of the normal traffic data (corresponding to various IoT clients) and use it to explore different autoencoder models. This exploration is not performed in a federated way. Each dataset is partitioned into 80\% training and 20\% evaluation. Among the tested models, we selected the simplest one (fewer parameters) that produced low enough evaluation loss to minimize overfitting problems.

Regarding the minimization of overfitting problems, the autoencoder training loss function includes a $L_2$ regularization term controlled by the $\lambda$ parameter, as noted in equation~\eqref{eq:loss_function}. The regularization directs the training in such a way as to make the model parameters smaller and prevent a single or few features from having too much weight in the model prediction results. While the autoencoder model selection step is performed in a non federated way, the final FL training process of the following steps can also have added advantages for preventing overfitting problems. According to McMahan et al.~\cite{McMahan2017}, one of the benefits of model averaging in FL is that it produces a regularization effect similar to the one achieved by dropout. In this case, FL helps to mitigate the overfitting problems that can occur in clients with fewer training data samples.

The number of nodes for the input and output layer of the autoencoder is fixed to the same number as the input feature dimensions (which can be 27 or 69, depending on the discretization method for the source and destination ports). For the encoder part, we evaluated different combinations with 1, 2 and 3~hidden layers, with each following layer having half as many nodes as the previous one~$\lfloor \frac{\text{\# nodes previous layer}}{2} \rfloor$, and a symmetric decoder. The selected autoencoder model and hyperparameters will be used in the next step: device clustering.

\subsection{Device Clustering}

Using the selected autoencoder model from the previous step, the device clustering process begins, which consists of the first phase detailed in Algorithm~\ref{alg:heterogeneousfl}.

The FL server initializes the selected autoencoder model and distributes it to all IoT clients (78 nodes in the city network). Each client locally trains the model for $\epsilon$~epochs using as the \textsc{ClientOpt} optimizer, the optimizer selected from the previous step. The partially trained models are uploaded back to the server to start the model fingerprinting and clustering process. As detailed in Algorithm~\ref{alg:heterogeneousfl}, the server flattens the parameters of each model and performs PCA to reduce the dimensionality of the parameters. We are going to select the number of components needed to explain at least 90\% of the variance. We use the K-means algorithm with the k-means++ initialization scheme to cluster the models, and hence the clients.

The experiments are repeated for different values of $\epsilon = 1, 2, 4, 8, 16$ and $32$, and the optimal number of clusters~$K$ is automatically selected based on the analysis of the following internal validation metrics: Silhouette, Davies-Bouldin and S\_Dbw.

\subsection{Federated Hyperparameter Tuning}

In this step, we are going to tune the rest of the FL hyperparameters. Each cluster identified in the previous step will have its own federated hyperparameter tuning. First, we are going to tune the \textsc{ClientOpt} and \textsc{ServerOpt} optimizer algorithms. Then, for the selected \textsc{ClientOpt} and \textsc{ServerOpt}, we are going to further refine the client and server learning rates. While in the previous step of autoencoder model selection the client optimizer and learning rates were selected, these values might not be optimal for the FL training process.

The \textsc{ClientOpt} and \textsc{ServerOpt} are tuned by comparing multiple combinations of SGD (with and without momentum) and Adam optimizers both for the clients and the server. SGD is tested without momentum and with momentum set to $0.9$ (as suggested in~\cite{Wang2021a}), for Adam two combinations are tested: $\beta_1 = 0.9$, $\beta_2 = 0.999$, $\epsilon = 1\times10^{-8}$ (default values defined in PyTorch) and $\beta_1 = 0.9$, $\beta_2 = 0.99$, $\epsilon = 0.001$ (as suggested in~\cite{Wang2021a}). In total, 16 combinations are evaluated. Each client trains on 80\% of its local data and is evaluated on the remaining 20\%; both losses are reported to the aggregation server. We select the optimizer combination that minimizes the average evaluation loss across all the cluster clients in fewer FL training rounds.

Then, the $\eta$ and $\eta_s$ learning rates are refined via grid search. For all the hyperparameter tuning we set $E = 1$, and use the same random seed to initialize the ML model parameters in order to reduce the effects of the random model initialization noise.

\subsection{Clustered Federated Learning}

After the device clustering and federated hyperparameter tuning steps, we perform the clustered federated learning using the fine-tuned \textsc{ClientOpt} and \textsc{ServerOpt} optimizers and their respective learning rates. We perform $R$ FL rounds to train the $K$ global models, one for each identified cluster. In this case, we repeat the process for different values of the number of local training epochs $E = 1, 2, 4$ and $8$ to evaluate its effect on the training process.

Similar to the previous steps, each client trains on 80\% of its local data and evaluates the model on the remaining 20\%. Each client records the loss for the training and evaluation splits after the local model training and sends it to the server. This is repeated for all the FL rounds. This way, the server can monitor both the training and evaluation loss progression and check if there are any overfitting signs.

\subsection{Anomaly Detection}

After training the models with FL, we are going to evaluate the anomaly detection performance of the resulting $K$ global models. Recall that at this step, each device holds a local copy of the global model that corresponds to its cluster. To estimate the unsupervised anomaly detection capabilities, we are going to use the validation-normal and validation-attack datasets.

\subsubsection{Threshold selection}

For each client, we will first evaluate the trained global model on its corresponding validation-normal dataset to estimate the anomaly detection threshold. Note that the evaluation is local; therefore, each device will compute its own threshold value. We will opt for a simplistic approach and select the largest MSE from the validation-normal dataset as the threshold. Packets with $\text{MSE} > \text{threshold}$ will be considered anomalous. Then, we will evaluate the same model on the multiple validation-attack datasets to identify all the anomalous packets.

\subsubsection{Anomaly detection performance}

The anomaly detection performance is measured by evaluating the trained global models on the multiple validation-attack datasets detailed in section~\ref{sec:data-generation-collection}. To obtain performance metrics, we will manually label the validation-attack datasets to provide ground truth labels to be compared with the detected anomalies from the autoencoder. The labeling process is based on the known IP addresses of the attacker, victim, IoT and cloud nodes, and attack timestamps extracted from the scenario. We recall that this ground truth labeling is only used to compute the performance metrics and is never used for training; also, the IP addresses are never used for model training. In a real deployment, prior labeling of the network data might not be feasible, but here it will give us an estimate of the performance of the global models to detect the attacks considered in our threat model; however, note that the manual labeling process is a heuristic and might misclassify some packets.

We provide the standard confusion matrix metrics: true positives (TP), false negatives (FN), false positives (FP), true negatives (TN) and their derivate metrics, including accuracy, F1 score and Matthews correlation coefficient (MCC).

\subsection{Baseline experimental comparisons}

We are considering Kitsune~\cite{Mirsky2018} network IDS and two non-clustered FL approaches for the baseline experimental comparisons.

\subsubsection{Kitsune}

Kitsune is a state-of-the-art network IDS that uses an ensemble of autoencoders trained in an unsupervised and online manner. The similarities of being unsupervised and based on autoencoders make it an interesting comparison; however, there are some fundamental differences between Kitsune and the proposed method. First, Kitsune does not use FL to train the model; it is deployed in each machine and only uses local data. Second, Kitsune uses features based on temporal statistics of network packets taken over multiple damped windows; instead, we extract features obtained from each packet in isolation. Third, Kitsune is trained in an online manner, so the training is performed using one sample at a time instead of multiple training iterations over batches of the data. Another difference is that Kitsune does not filter IPv6 or ARP packets.

\subsubsection{non-clustered FL with weighted aggregation}

This FL baseline is identical to the proposed approach from Algorithm~\ref{alg:heterogeneousfl}, except that the clustering step is removed. That is, we consider $K = 1$ (all clients belong to the same cluster), and the objective is to train a single global model that fits all the clients in the federated network.

\subsubsection{non-clustered FL without weighted aggregation}

In the server model aggregation step, the contribution of each client is weighted by the number of training samples used by that client (Algorithm~\ref{alg:fl}). This process can bias the global model towards clients with more training samples. Hence, this FL baseline is the same as the previous non-clustered FL baseline ($K = 1$, single global model for all clients), except that at the server aggregation step, the contribution of all clients will be equally weighted.

Both non-clustered FL baselines are used to experimentally compare whether clustering offers significant advantages for unsupervised anomaly detection in FL settings.

\section{Results} \label{sec:results}

In this section, we present the results obtained from the experiments described in section~\ref{sec:implementation}.

\subsection{Autoencoder Model Selection}

As previously stated, the input and output shapes of the autoencoder are the same as the number of input data feature dimensions. Depending on the selected source and destination port discretization method from the network data processing step, the dimensions are 27 or 69 for the three-range and hierarchical discretization, respectively. For the autoencoder model selection, we will consider both cases. The distinction between the two methods will be shown later in the device clustering results.

We detected no significant improvement in the validation loss after 2~hidden encoder layers, irrespective of the port discretization method. Thus, for the three-range discretization method, the final autoencoder model is a two hidden layer encoder with 13~and~6 nodes, respectively, and a symmetric decoder with 6~and~13 nodes. For the hierarchical discretization, the encoder layers include 34~and~17 nodes with a symmetric decoder. We use the $ReLU$~activation function after each layer. The optimizer is Adam with a $1\times10^{-3}$ learning rate, $L_2$ regularization weight from equation~\eqref{eq:loss_function} $\lambda = 1\times10^{-5}$ and a batch size of~32.

\subsection{Device Clustering}

\begin{figure*}
	\centering
	
	\subfloat[Unsupervised clustering quality scores.\label{fig:clus_4e_3p_unsup}]{%
		\includegraphics[width=0.32\linewidth]{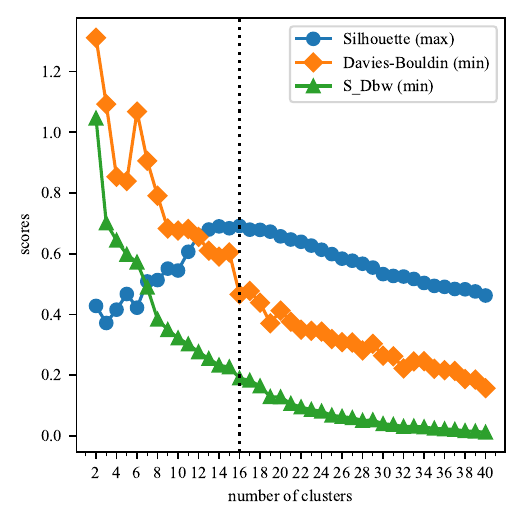}
	}
	\hfill
	\subfloat[Clustering quality scores compared to ground truth labeling.\label{fig:clus_4e_3p_sup}]{%
		\includegraphics[width=0.32\linewidth]{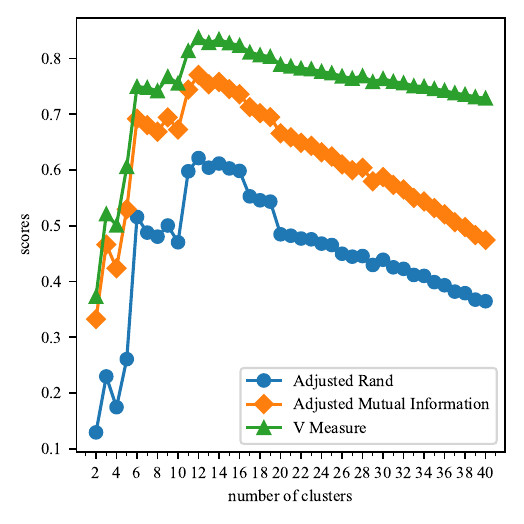}
	}
	\hfill
	\subfloat[2D projection of the model fingerprints clustered into $K = 16$ groups.\label{fig:clus_4e_3p_pca}]{%
		\includegraphics[width=0.32\linewidth]{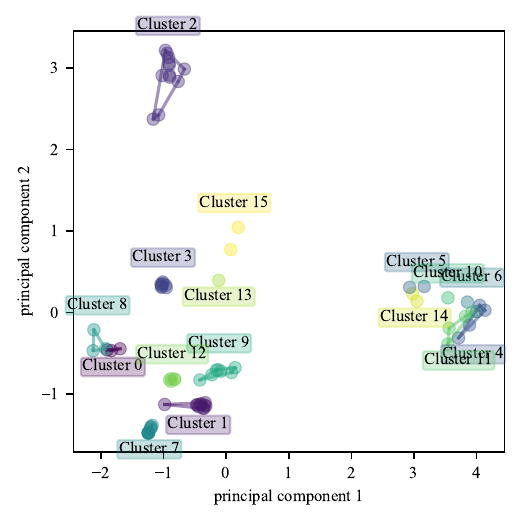}
	}
	
	\caption{Device clustering results for $\epsilon = 4$ using the three-range discretization strategy for the source and destination ports.}
	\label{fig:clus_4e_port_three}
\end{figure*}

\begin{figure*}
	\centering
	
	\subfloat[Unsupervised clustering quality scores.\label{fig:clus_4e_hp_unsup}]{%
		\includegraphics[width=0.32\linewidth]{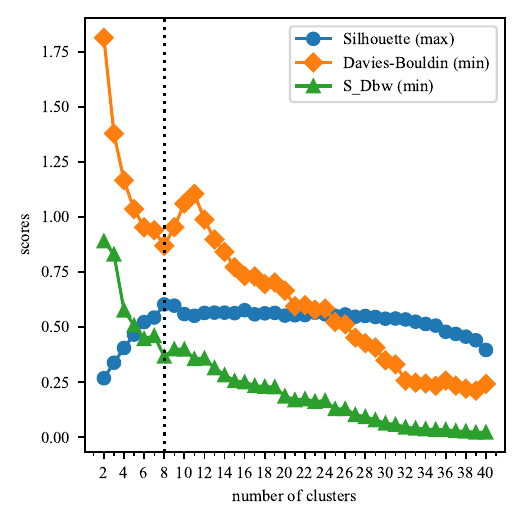}
	}
	\hfill
	\subfloat[Clustering quality scores compared to ground truth labeling.\label{fig:clus_4e_hp_sup}]{%
		\includegraphics[width=0.32\linewidth]{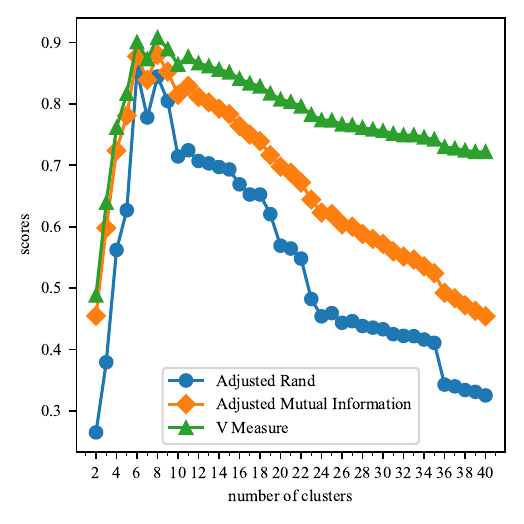}
	}
	\hfill
	\subfloat[2D projection of the model fingerprints clustered into $K = 8$ groups.\label{fig:clus_4e_hp_pca}]{%
		\includegraphics[width=0.32\linewidth]{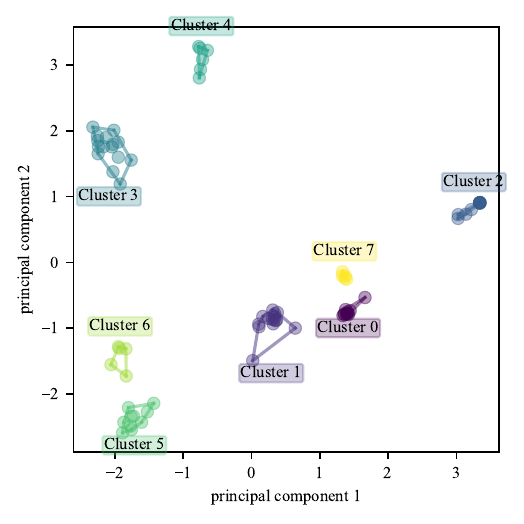}
	}
	
	\caption{Device clustering results for $\epsilon = 4$ using the hierarchical discretization strategy for the source and destination ports.}
	\label{fig:clus_4e_port_hier}
\end{figure*}

The clustering experiments are repeated for the two port discretization methods and multiple values of the local training epochs $\epsilon = 1, 2, 4, 8, 16$ and $32$ using the client optimizer parameters obtained from the previous autoencoder model selection step. For each value of $\epsilon$, to identify the optimal number of clusters~$K$, we perform K-means clustering with $K$ ranging from $2$ to $40$ cluster centroids. The results for $\epsilon = 4$ using the three-range port discretization method are shown in \figurename~\ref{fig:clus_4e_port_three}, while the results for $\epsilon = 4$ using the hierarchical discretization method are shown in \figurename~\ref{fig:clus_4e_port_hier}.

The unsupervised clustering quality scores are shown in \figurename~\ref{fig:clus_4e_3p_unsup} and \figurename~\ref{fig:clus_4e_hp_unsup}. For Silhouette higher scores represent better clusters, for Davies-Bouldin and S\_Dbw lower scores represent better clusters. The dotted vertical line marks the selected $K$ for each discretization method. For the three-range discretization, the Silhouette score is maximized at $K=16$, and the Davies-Bouldin score shows a dip at the same point. For S\_Dbw the score is monotonously decreasing; however, 16 is a good number of clusters based on the elbow method. For the hierarchical discretization method, the Silhouette score is maximized at $K=8$ and both Davies-Bouldin and S\_Dbw show a dip at that point. A 2D projection of the model fingerprints and the clustering results is represented in \figurename~\ref{fig:clus_4e_3p_pca} with $K=16$ and \figurename~\ref{fig:clus_4e_hp_pca} with $K=8$.

Due to the differences in the clustering results depending on the port discretization strategy, we perform an additional experiment. For each strategy, we measure the similarity between the K-means clustering labels and the ground truth clustering using the adjusted Rand index, adjusted mutual information score and the V-measure score. To create the ground truth labeling, we assign each instance a label based on its template type according to the data in Table~\ref{tab:iot_device_templates} (12 different labels). The results are shown in \figurename~\ref{fig:clus_4e_3p_sup} and \figurename~\ref{fig:clus_4e_hp_sup}. For the three-port discretization, the score is maximized when $K=12$ (in contrast to the $K=16$ from the internal validation metrics). The results for the hierarchical discretization method in \figurename~\ref{fig:clus_4e_hp_sup} show a maximum in $K=8$ (same results as with the internal validation metrics) and overall higher scores compared to the previous method.

Using the hierarchical discretization method, we obtain an agreement in the optimal value for $K$ between the unsupervised clustering and the similarity with the ground truth scores. It also creates a clearer distinction between the clusters (\figurename~\ref{fig:clus_4e_hp_pca}) compared to the three-port discretization method (\figurename~\ref{fig:clus_4e_3p_pca}). Additionally, it may be desirable to lean towards small $K$ values so that the FL process benefits from a larger cohort size for each cluster. From now on, we are going to use the hierarchical discretization method and $K=8$ for the rest of the experimentation.

In real deployment settings, where there might be no ground truth labels for the device types, only unsupervised internal clustering validation metrics will be available to analyze the clustering quality. Hence, from the experimental results, we infer that in order to select the number of clusters $K$, a robust approach is first to consider the value that maximizes the Silhouette score. In cases where there are different values of $K$ with similar scores, break ties by considering the Davies-Bouldin and S\_Dbw metrics.

The device clustering results from \figurename~\ref{fig:clus_4e_hp_pca} are shown in Table~\ref{tab:clustering_results}. All the IP camera related devices are grouped into the same cluster. Interestingly, for devices of the same type, the clustering method can distinguish between those communicating via plain text or over an encrypted channel; for example, the Predictive maintenance devices in Cluster 5 and Cluster 6 or the Combined cycle in Cluster 2 and Cluster 7. Clusters 0, 2 and 4 are composed of heterogeneous devices; however, the devices in the same cluster communicate using the same primary protocol: MQTT, CoAP and RTSP, respectively.

Regarding the clustering results for the other tested values of $\epsilon$, the results in the $\epsilon = 8$ case are very similar to the discussed $\epsilon = 4$ case, where the unsupervised and supervised clustering validation metrics agree on the optimal number of groups. However, in some cases, the number of clusters decreases to 7, merging the groups of the same device types that communicate in plain or over an encrypted channel. In the $\epsilon = 2$ case, the number of clusters according to the unsupervised metrics is 9, and in the $\epsilon = 1$ case, it is increased to 11. Both cases tend to split the groups formed by the IP camera devices and Predictive maintenance ones. For $\epsilon = 16$ and 32, the number of identified clusters using unsupervised metrics also tends to increase to around 11 and 17, respectively; moreover, for both cases, the supervised metrics still show the optimum at 8, indicating a discrepancy between the unsupervised and supervised metrics for higher values of $\epsilon$.

\begin{table}[t]
	\small
	\centering
	
	\caption{Unsupervised clustering results using the hierarchical port discretization strategy for $\epsilon = 4$ and $K = 8$.}
	\label{tab:clustering_results}
	
	\begin{tabularx}{1.0\linewidth}{l|X}
		\toprule
		Cluster name   & Cluster contents                                                      \\
		\midrule
		Cluster 0      & Air quality (x1), Building monitor (x5), Domotic monitor (x5)         \\
		Cluster 1      & Hydraulic system (x15)                                                \\
		Cluster 2      & City power (x1), Combined cycle (x10)                                 \\
		Cluster 3      & Cooler motor (x15)                                                    \\
		Cluster 4      & IP camera museum (x2), IP camera street (x2), IP camera consumer (x2) \\
		Cluster 5      & Predictive maintenance (x10)                                          \\
		Cluster 6      & Predictive maintenance (x5)                                         \\
		Cluster 7      & Combined cycle tls (x5)                                               \\
		\bottomrule
	\end{tabularx}
	
\end{table}

From \figurename~\ref{fig:clus_4e_hp_pca}, while some clusters are clearly separated from the rest, others, such as clusters 0, 1 and 7 or clusters 6 and 5, seem to be close in the 2D projection. The dimensionality of the model fingerprints is the same as the number of PCA components needed to explain at least 90\% of the model parameter variance, which in this particular case is 23. \figurename~\ref{fig:clus_4e_hp_pca} only shows the first two dimensions, corresponding to approximately 40\% of the explained variance. This might indicate that clusters that are close to each other in the 2D representation are also close in the higher-dimensional space. When the training data amount of each device is insufficient or due to the random model initialization influence, the clustering results' stability might be affected for those groups close to each other. In order to study the clustering stability, we are going to perform an additional experiment.

\subsubsection{Cluster stability for varying training data size}

In this experiment, we will repeat the device clustering process for the $\epsilon = 4$ and the hierarchical discretization port method case. However, we will vary the fraction of training data used to partially train the models. The server initializes an autoencoder model, and each client will randomly subsample a fraction of its own local training data. The clustering procedure is the same as explained before, only that each client performs local training for $\epsilon$ epochs using only the specified fraction of the data. A different experiment is conducted for the following fractions: 1\%, 10\%, 20\%, 30\%, 40\%, 50\%, 60\%, 70\%, 80\%, 90\% and 99\%. All those experiments are going to be further repeated 30 times to account for any variability in the results due to the random model initialization effect at the server and the random subsampling process at each device.

\begin{figure}
	\centering
	\includegraphics[width=\linewidth]{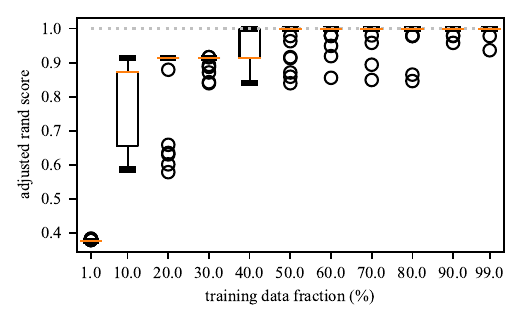}
	\caption{Cluster stability results for varying training data sizes. Each training data fraction percentage shows a boxplot for 30 repetitions of the adjusted Rand score with the clustering results from Table~\ref{tab:clustering_results} ($K=8$). Circles indicate outliers, i.e., samples outside 1.5x of the inter-quartile range. The orange line in each boxplot shows the medians.}
	\label{fig:clus_stability}
\end{figure}

To measure the cluster stability for varying training data sizes, we select $K=8$ and compare the clustering results of the new experiments with the clustering results obtained in Table~\ref{tab:clustering_results}. The similarity is measured using the adjusted Rand score. A value of $1.0$ is obtained when the clusterings are identical, and values near $0.0$ indicate random labeling. The results are shown in \figurename~\ref{fig:clus_stability}. For training data fractions $\ge 50\%$, the majority of runs achieve a score of $1.0$, showing that the clustering results are mostly stable, but some outliers appear. The number of outliers is reduced with increasing training data fraction. When the training data fraction is reduced below 50\%, the clustering quality is negatively affected.

We also note that for small training data fractions, the optimal value of $K$ shown by unsupervised internal clustering validation metrics tends to decrease, is more unstable from repetition to repetition and diverges considerably from the optimal value shown by supervised clustering metrics using the ground truth labeling. The contrary occurs when the training data fraction is $\ge 60\%$, where the unsupervised and supervised metrics are close, and $K$ is around $8 \pm 1$.

\subsection{Federated Hyperparameter Tuning}

To select the \textsc{ClientOpt} and \textsc{ServerOpt} optimizer algorithms, for each cluster we performed 16~trials consisting of different combinations of SGD and Adam as defined in Table~\ref{tab:hyperparameter_trials}. Client learning rates are fixed to $\eta = 1\times10^{-3}$ and the $L_2$ regularization weight from equation~\eqref{eq:loss_function} is set to $\lambda = 1\times10^{-5}$.

\begin{figure*}
	\centering
	
	\subfloat[Cluster 0 (MQTT).\label{fig:opt_clus0}]{%
		\includegraphics[width=0.32\linewidth]{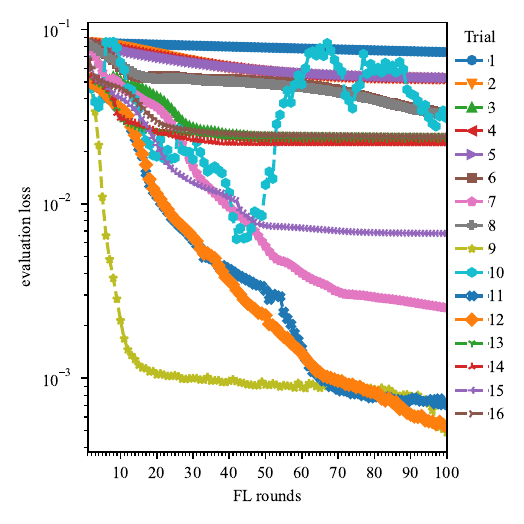}
	}
	\hfill
	\subfloat[Cluster 2 (CoAP).\label{fig:opt_clus2}]{%
		\includegraphics[width=0.32\linewidth]{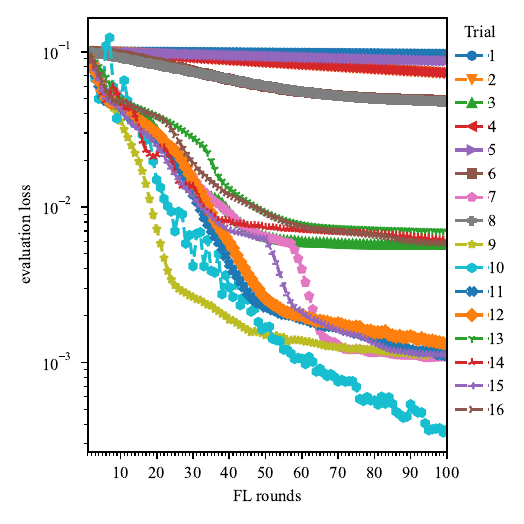}
	}
	\hfill
	\subfloat[Cluster 4 (Camera).\label{fig:opt_clus4}]{%
		\includegraphics[width=0.32\linewidth]{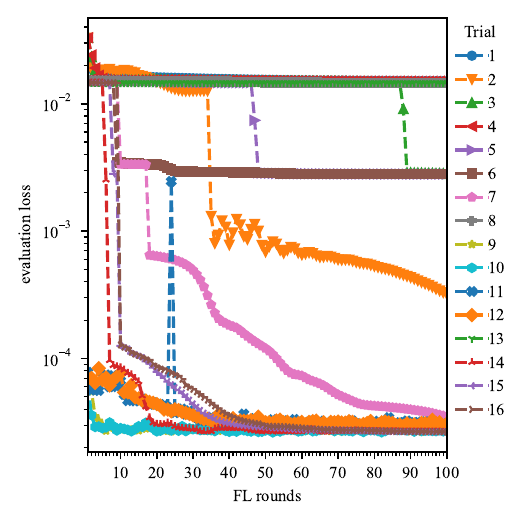}
	}
	
	\caption{Federated hyperparameter tuning, \textsc{ClientOpt} and \textsc{ServerOpt} optimizer selection. $E = 1$. The trials are defined in Table~\ref{tab:hyperparameter_trials}.}
	\label{fig:clientopt_serveropt}
\end{figure*}

We show the results of the mean evaluation loss progression for 100~FL rounds and all the trials for Clusters 0, 2 and 4 in \figurename~\ref{fig:clientopt_serveropt}, as these three clusters are more complex than the others because they are formed by heterogeneous devices. In general, including adaptive optimization methods for \textsc{ClientOpt}, \textsc{ServerOpt} or both provides faster convergence and smaller losses compared to the standard SGD; however, some combinations have difficulty to converge, showing an increasing loss trend as in Trial~10 for Cluster 0 (\figurename~\ref{fig:opt_clus0}). For Cluster 0, Trial 9 clearly shows faster convergence speeds and a smaller evaluation loss after 100~FL rounds. Trial 12 also shows a similar evaluation loss at the last round, but at a much slower convergence rate. For Clusters 2 and 4 (\figurename~\ref{fig:opt_clus2} and \figurename~\ref{fig:opt_clus4}), Trials 9 and 10 show the best performance. Trial 10 from Cluster 2 reaches a smaller loss than Trial 9; however, by fine tuning the Trial 9 learning rates, it can reach the same loss values.

\begin{figure*}
	\centering
	
	\subfloat[Cluster 0 (MQTT). Grid search for Trial 9.\label{fig:gs_clus0}]{%
		\includegraphics[width=0.32\linewidth]{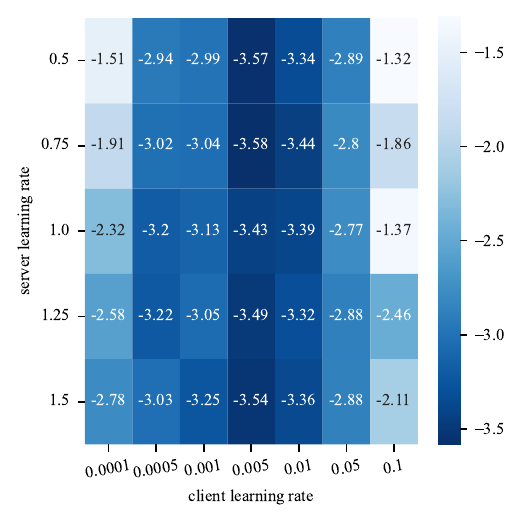}
	}
	\hfill
	\subfloat[Cluster 2 (CoAP). Grid search for Trial 9.\label{fig:gs_clus2}]{%
		\includegraphics[width=0.32\linewidth]{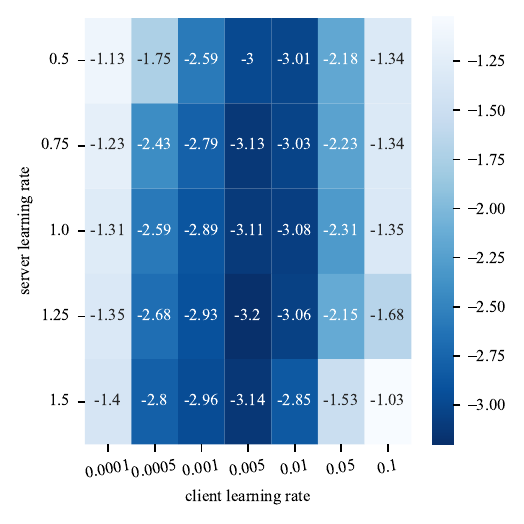}
	}
	\hfill
	\subfloat[Cluster 4 (Camera). Grid search for Trial 9.\label{fig:gs_clus4}]{%
		\includegraphics[width=0.32\linewidth]{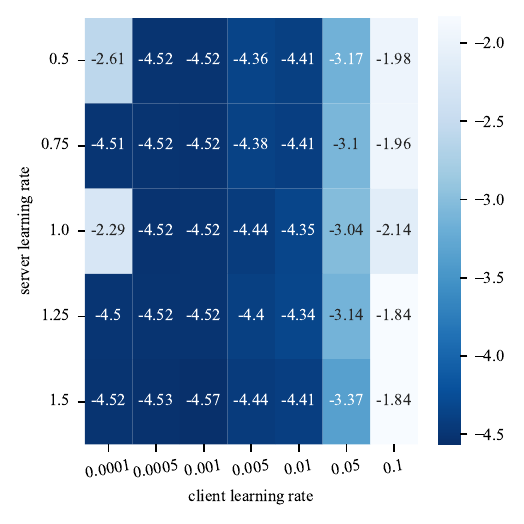}
	}
	
	\caption{Federated hyperparameter tuning, $\eta$ and $\eta_s$ learning rate grid search. The values represent base~10 logarithm of the evaluation loss after 60~FL rounds.}
	\label{fig:gridsearch}
\end{figure*}

The client and server learning rates ($\eta$ and $\eta_s$, respectively) for each cluster are fine tuned by performing a grid search varying both values simultaneously. The results are shown in the heat maps from \figurename~\ref{fig:gridsearch}. The heat maps show the logarithm of the evaluation loss after 60~FL rounds; darker colors show a smaller loss. For the three cases, many combinations achieve a similar low loss; we are going to select the combination with a smaller loss for all cases.

The final optimizer selection are as follows. Cluster 0 \textsc{ClientOpt} is Adam1 with $\eta = 0.005$, and \textsc{ServerOpt} is SGD with $\eta_s = 0.75$. Cluster 2 \textsc{ClientOpt} is Adam1 with $\eta = 0.005$,  and \textsc{ServerOpt} is SGD with $\eta_s = 1.25$. Cluster 4 \textsc{ClientOpt} is Adam1 with $\eta = 0.001$, and \textsc{ServerOpt} is SGD with $\eta_s = 1.5$.

\begin{table}[t]
	\small
	\centering

	\caption{\textsc{ClientOpt} and \textsc{ServerOpt} combinations for each hyperparameter tuning trial. ``SGD'' is SGD without momentum, ``SGDm'' refers to SGD with momentum $0.9$, ``Adam1'' refers to $\beta_1 = 0.9$, $\beta_2 = 0.999$, $\epsilon = 1\times10^{-8}$ and ``Adam2'' refers to Adam $\beta_1=0.9$, $\beta_2=0.99$, $\epsilon=10^{-3}$. $\lambda$ is set to $1\times10^{-5}$ for all trials.}
	\label{tab:hyperparameter_trials}

	\begin{tabularx}{1.0\linewidth}{l|Xl|Xl}
		\toprule
		Trial    & \textsc{ClientOpt} & $\eta$           & \textsc{ServerOpt} & $\eta_s$ \\
		\midrule
		Trial 1  & SGD                & $1\times10^{-3}$ & SGD                & $1.0$      \\
		Trial 2  & SGD                & $1\times10^{-3}$ & SGDm               & $1.0$      \\
		Trial 3  & SGD                & $1\times10^{-3}$ & Adam1              & $1\times10^{-2}$      \\
		Trial 4  & SGD                & $1\times10^{-3}$ & Adam2              & $1\times10^{-2}$      \\
		Trial 5  & SGDm               & $1\times10^{-3}$ & SGD                & $1.0$      \\
		Trial 6  & SGDm               & $1\times10^{-3}$ & SGDm               & $1.0$      \\
		Trial 7  & SGDm               & $1\times10^{-3}$ & Adam1              & $1\times10^{-2}$      \\
		Trial 8  & SGDm               & $1\times10^{-3}$ & Adam2              & $1\times10^{-2}$      \\
		Trial 9  & Adam1              & $1\times10^{-3}$ & SGD                & $1.0$      \\
		Trial 10 & Adam1              & $1\times10^{-3}$ & SGDm               & $1.0$      \\
		Trial 11 & Adam1              & $1\times10^{-3}$ & Adam1              & $1\times10^{-2}$      \\
		Trial 12 & Adam1              & $1\times10^{-3}$ & Adam2              & $1\times10^{-2}$      \\
		Trial 13 & Adam2              & $1\times10^{-3}$ & SGD                & $1.0$      \\
		Trial 14 & Adam2              & $1\times10^{-3}$ & SGDm               & $1.0$      \\
		Trial 15 & Adam2              & $1\times10^{-3}$ & Adam1              & $1\times10^{-2}$      \\
		Trial 16 & Adam2              & $1\times10^{-3}$ & Adam2              & $1\times10^{-2}$      \\
		\bottomrule
	\end{tabularx}
\end{table}

\subsection{Clustered Federated Learning}

\begin{figure*}
	\centering
	
	\subfloat[Cluster 0 (MQTT).\label{fig:boxplot_clus0}]{%
		\includegraphics[width=0.32\linewidth]{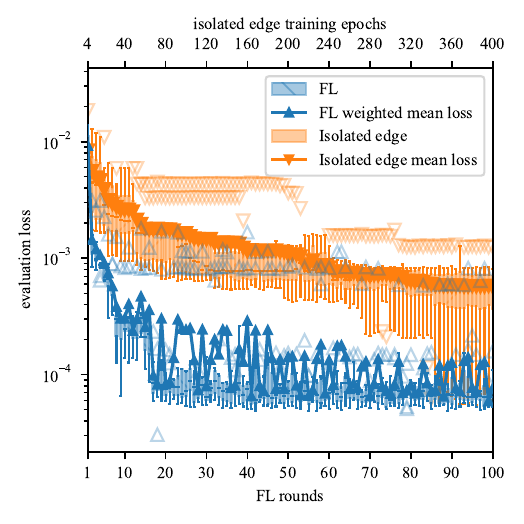}
	}
	\hfill
	\subfloat[Cluster 2 (CoAP).\label{fig:boxplot_clus2}]{%
		\includegraphics[width=0.32\linewidth]{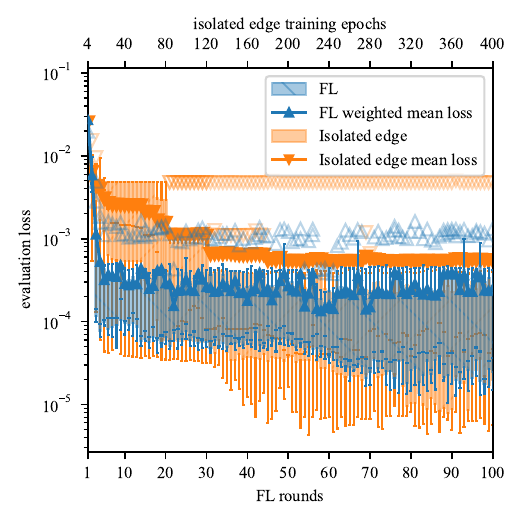}
	}
	\hfill
	\subfloat[Cluster 4 (Camera).\label{fig:boxplot_clus4}]{%
		\includegraphics[width=0.32\linewidth]{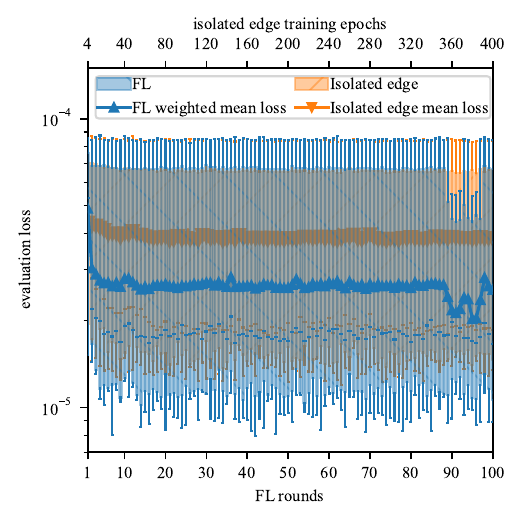}
	}
	
	\caption{Clustered FL training progression for $E=4$ local training epochs and $R=100$ FL rounds (blue boxplots). It is compared with isolated edge training where each device trains on its own dataset for $R \times E = 400$ epochs (orange boxplots).}
	\label{fig:boxplot_fl_nofl}
\end{figure*}

The final FL training process is performed using the client-side and server-side optimizers and learning rates obtained after the federated hyperparameter tuning described in the previous step for each identified cluster. We repeated the experiments for different values of the number of local training epochs $E = 1, 2, 4$ and 8. Increasing the number of local training epochs generally leads to lower loss values and fewer FL rounds to reach convergence at the expense of more local computation time. However, we also observed an increased variance in the loss distribution across the devices of the cluster when using large numbers of local training epochs. The training results for $E=4$ local epochs and $R=100$ FL rounds are shown in \figurename~\ref{fig:boxplot_fl_nofl} for Clusters 0, 2 and 4. Each boxplot shows the evaluation loss distribution across the devices of the cluster at a certain FL round.

The progression of both training and evaluation losses was checked, there was a small gap between the training and evaluation loss, however, this gap remained more or less constant for all the FL rounds and did not show overfitting patterns.

\subsubsection{Training baseline}

As a training baseline, we performed additional experiments to compare the training evaluation loss progression between FL and isolated edge training, where each device trains on its local data without cooperation. In isolated training, each device in the cluster starts with a random initialization of the autoencoder and trains it using the same client-side optimizer as in the FL case. The training is performed for a total of $R \times E$~epochs so that the amount of local training performed by each device is comparable to the FL case. The comparison is shown in \figurename~\ref{fig:boxplot_fl_nofl}.

For Clusters 0 and 2, there is a noticeable gap in the evaluation loss between the FL and isolated training methods, where FL shows a faster convergence rate, especially in early rounds. For Cluster 4, while FL shows a lower average loss, the loss distribution is similar to the isolated training.

This difference might be explained due to the different training data volumes generated by each device. Cluster 4 devices generate a much larger data volume because they are comprised of image streaming devices, ranging between 300 to 800 MB of raw pcap data; this extensive training data can benefit local isolated training. However, the raw pcap data for Cluster 0 devices ranges approximately between 230 KB to 270 KB. For Cluster 2 devices, the raw data is between 100 KB to 170 KB. This suggests the advantages of using FL for devices that generate a low volume of training data samples.

\subsection{Anomaly Detection}

Here we provide the anomaly detection performance results for clusters 0, 2 and 4 by evaluating the trained global models from the previous step on the multiple validation-attack datasets described in section~\ref{sec:data-generation-collection}. The number of packets (normal and attack) after filtering the pcaps is shown in table~\ref{tab:valid-attack-numpackets}. As explained in section~\ref{sec:implementation}, the anomaly threshold of each device is selected so that there are no false positives in the device's validation-normal dataset. The attack packets are considered as the positive class.

\begin{table}[t]
	\small
	\centering
	
	\caption{The number of packets after the IPv6 and ARP filtering step in the validation-attack datasets.}
	\label{tab:valid-attack-numpackets}
	
	\begin{tabularx}{1.0\linewidth}{X|l|l|l}
		\toprule
		validation-attack-* & Cluster 0 & Cluster 2 & Cluster 4 \\ \midrule
		mirai-scan-load     & 110,354   & 70,592    & 888,833   \\
		mirai-cnc-dos       & 6,810,612 & 7,777,427 & 7,924,688 \\
		merlin-cnc-dos      & 32,282    & 31,014    & 868,586   \\
		masscan             & 816       & 571       & 203,814   \\
		scan-amplification  & n/a       & 68,422    & n/a       \\ \bottomrule
	\end{tabularx}
\end{table}

\subsubsection{Cluster 0 (MQTT)}

We evaluate the global model of Cluster 0 on the four validation-attack datasets captured from one instance of the Building monitor device.

For the mirai-scan-load data, the reconstruction error of all anomalous packets is above the threshold, and the normal packets have a low reconstruction error except for a single false positive: TP, FN, FP, TN = 108532, 0, 1, 1821 (0.9999 accuracy, 0.9999 F1 and 0.9997 MCC). Similarly, for the masscan data the model correctly classified all packets: TP, FN, FP, TN = 528, 0, 0, 288 (1.0 accuracy, F1 and MCC).

For the mirai-cnc-dos and merlin-cnc-dos datasets, some false negatives are reported. In the mirai-cnc-dos case, the C\&C activity and seven out of eight DoS attacks were all correctly classified; however, some (but not all) of the attack packets corresponding to the same time frame when the DNS attack was being performed were below the anomaly threshold: TP, FN, FP, TN = 6743222, 66190, 0, 1200 (0.9903 accuracy, 0.9951 F1 and 0.1328 MCC). For the merlin-cnc-dos case, the C\&C activity, ingress tool transfer and three out of four attacks were all correctly classified. The model did not detect the anomalous packets corresponding to the ICMP flood attack: TP, FN, FP, TN = 25828, 5277, 0, 1177 (0.8365 accuracy, 0.9073 F1 and 0.3891 MCC). The reconstruction error scatter plot for the merlin-cnc-dos case is shown in \figurename~\ref{fig:anomaly_flids_c0_merlin}.

\subsubsection{Cluster 2 (CoAP)}

The global model of Cluster 2 is evaluated on the five validation-attack datasets captured from one instance of the Combined cycle device.

This model correctly classified all normal and anomalous packets for all the validation-attack datasets except for a single false negative packet. The mirai-scan-load case obtained: TP, FN, FP, TN =  70195, 0, 0, 397 (1.0 accuracy, F1 and MCC). For the mirai-cnc-dos data: TP, FN, FP, TN =  7777173, 0, 0, 254 (1.0 accuracy, F1 and MCC). In the merlin-cnc-dos case: TP, FN, FP, TN =  30754, 0, 0, 260 (1.0 accuracy, F1 and MCC). The masscan data obtained: TP, FN, FP, TN =  522, 0, 0, 49 (1.0 accuracy, F1 and MCC). And lastly, the scan-amplification: TP, FN, FP, TN =  68237, 1, 0, 184 (0.9999 accuracy, 0.9999 F1 and 0.9973 MCC). The reconstruction error for the scan-amplification dataset is shown in \figurename~\ref{fig:anomaly_flids_c2_nmapampli}.

\subsubsection{Cluster 4 (Camera)}

We evaluate the global model of Cluster 4 on the four validation-attack datasets captured from one instance of the IP camera museum.

\begin{figure*}
	\centering
	
	\subfloat[Cluster 0 global model evaluated on validation-attack-merlin-cnc-dos dataset for the same cluster. (A) start Merlin agent. (B) hping3 upload. (C), (D), (E) and (F) ICMP, UDP, SYN and ACK attacks, respectively. (G) stop Merlin agent. The (C) event packets are below the threshold.\label{fig:anomaly_flids_c0_merlin}]{%
		\includegraphics[width=0.32\linewidth]{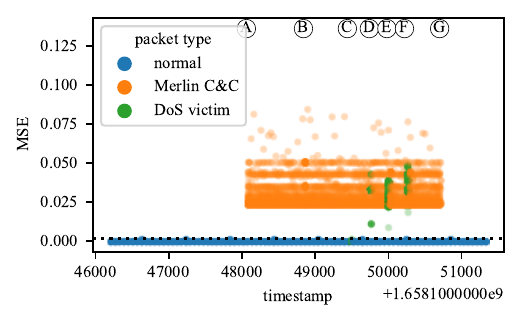}
	}
	\hfill
	\subfloat[Cluster 2 global model evaluated on validation-attack-scan-amplification dataset for the same cluster. (A)--(B) Nmap random port scan. (C) Nmap port 5683 scan. (D)--(E) Nmap top 1000 ports scan. (F) CoAP amplification attack.\label{fig:anomaly_flids_c2_nmapampli}]{%
		\includegraphics[width=0.32\linewidth]{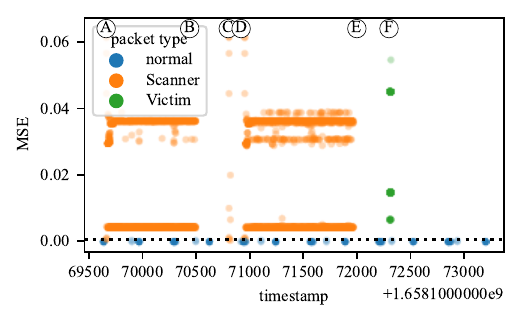}
	}
	\hfill
	\subfloat[Cluster 4 global model evaluated on validation-attack-mirai-scan-load dataset for the same cluster. (A) testbed's Mirai bot is started. (B) This Cluster 4 device (IP camera museum) gets infected with Mirai. Packets labeled as 'Others' are other Mirai-infected IoT devices scanning the network.\label{fig:anomaly_flids_c4_miraiscanload}]{%
		\includegraphics[width=0.32\linewidth]{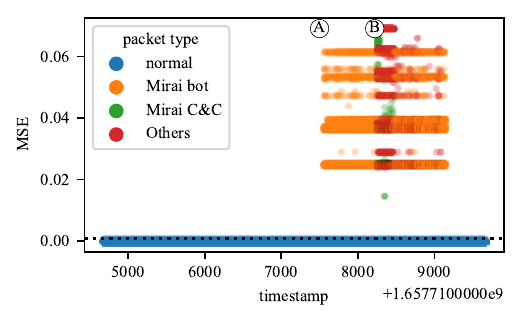}
	}
	
	\caption{Anomaly detection examples. Dotted line indicates the anomaly threshold, packets with MSE above the threshold are considered anomalous.}
	\label{fig:anomaly_flids}
\end{figure*}

For the mirai-scan-load data, the reconstruction error of all anomalous packets is above the threshold, and the normal packets are correctly classified except for a false positive: TP, FN, FP, TN =  81604, 0, 1, 807228 (0.9999 accuracy, 0.9999 F1 and 0.9999 MCC). This case is shown in \figurename~\ref{fig:anomaly_flids_c4_miraiscanload}. The number of false positives and false negatives is slightly increased in the mirai-cnc-dos dataset, part (but not all) of the packets corresponding to the time frame where the Mirai GRE IP and GRE Ethernet attacks are below the threshold: TP, FN, FP, TN =  7424929, 224, 4, 499531 (0.9999 accuracy, 0.9999 F1 and 0.9997 MCC).

For the merlin-cnc-dos dataset, all the packets were correctly classified: TP, FN, FP, TN =  30990, 0, 0, 837596 (1.0 accuracy, F1 and MCC). Similarly, the packets of the masscan dataset were also correctly classified: TP, FN, FP, TN =  548, 0, 0, 203266 (1.0 accuracy, F1 and MCC).

\subsection{Baseline experimental comparisons}

Here we provide anomaly detection performance results for the considered baseline approaches.

\subsubsection{Kitsune}

For the comparison with Kitsune, we use its publicly available Python implementation~\cite{KitsuneRepo}. We configure Kitsune to use the default parameters ($m = 10$ maximum size for any autoencoder in the ensemble layer). Kitsune does not use FL, so for each device on which we deploy it, we use its corresponding normal traffic data for training (the first 10\% to learn Kitsune's feature mapping and the remaining 90\% for the training of the autoencoder ensemble itself). Then, it is evaluated on the device's corresponding validation-attack datasets. In this comparison experiment, we are not primarily interested in the results of the anomaly detection metrics; however, we are interested in what kind of attacks or malicious behavior detection our proposed method differs from Kitsune.

Regarding the mirai-scan-load dataset, the measured metrics ranged from 0.9729 -- 0.9771 accuracy, 0.9861 -- 0.9883 F1 and 0.3518 -- 0.4525 MCC depending on the device type. Overall, most packets related to the Mirai scanning, brute-forcing and malware loading stages appeared above the anomaly threshold. However, most Mirai C\&C related traffic went undetected.

\begin{figure*}
	\centering
	
	\subfloat[Proposed clustered FL method.\label{fig:flids_vs_kits_miraidos_flids}]{%
		\includegraphics[width=0.49\linewidth]{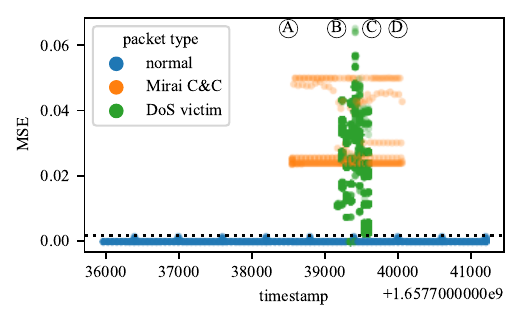}
	}
	\hfill
	\subfloat[Kitsune.\label{fig:flids_vs_kits_miraidos_kits}]{%
		\includegraphics[width=0.49\linewidth]{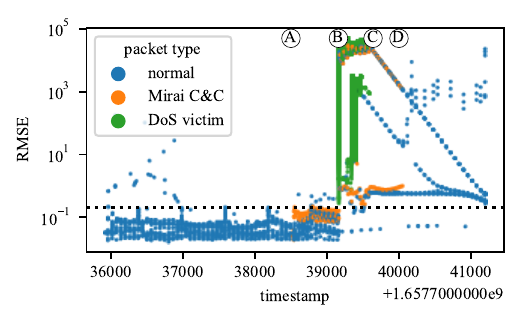}
	}
	
	\caption{Anomaly detection results for the validation-attack-mirai-cnc-dos dataset on one of the Building monitor devices. (A) Start Mirai bot on the device. (B)--(C) DoS attacks. (D) stop Mirai bot. Dotted line indicates the anomaly threshold.}
	\label{fig:flids_vs_kits_miraidos}
\end{figure*}

The results on the Mirai C\&C related traffic are best observed on the mirai-cnc-dos datasets, shown in \figurename~\ref{fig:flids_vs_kits_miraidos_kits}. The measured metrics are 0.9998 accuracy, 0.9999 F1 and 0.7397 MCC. Kitsune correctly detected all the performed DoS attacks as anomalous, but failed to detect the C\&C related traffic. During the period between the Mirai bot activation and the first attack, the device periodically communicates with the Mirai C\&C server. This traffic went undetected for Kitsune as its reconstruction error is close to the error for normal traffic. In contrast, while our proposed method failed to detect some packets related to the DoS attacks, all the Mirai C\&C traffic is well separated from the normal activity, as shown in \figurename~\ref{fig:flids_vs_kits_miraidos_flids}.

\begin{figure*}
	\centering
	
	\subfloat[Proposed clustered FL method.\label{fig:flids_vs_kits_masscan_flids}]{%
		\includegraphics[width=0.49\linewidth]{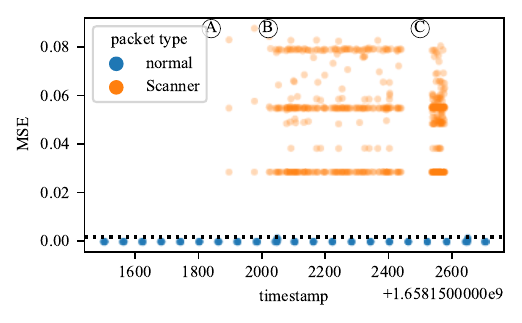}
	}
	\hfill
	\subfloat[Kitsune.\label{fig:flids_vs_kits_masscan_kits}]{%
		\includegraphics[width=0.49\linewidth]{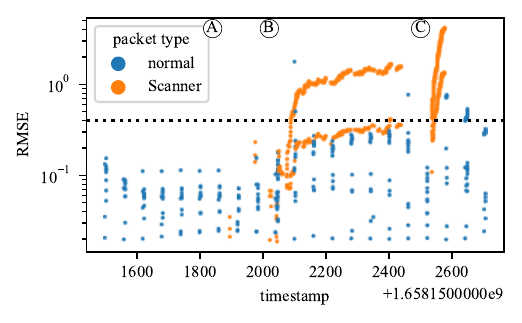}
	}
	
	\caption{Anomaly detection results for the validation-attack-masscan dataset on one of the Building monitor devices. (A) Masscan node performs low-volume scan. (B) Masscan node performs medium-volume scan. (C) Masscan node performs high volume scan. Dotted line indicates the anomaly threshold.}
	\label{fig:flids_vs_kits_masscan}
\end{figure*}

The masscan dataset also shows significant differences between Kitsune and the proposed clustered FL model. The measured metrics ranged from 0.7781-0.8122 accuracy, 0.8026-0.8865 F1 and 0.4665-0.5971 MCC. All the low-volume scanning activity and a significant number of packets from the medium-volume scanning activity were below Kitsune's threshold, as shown in \figurename~\ref{fig:flids_vs_kits_masscan_kits}. However, our proposed method detected all activity irrespective of the scanning rate, as shown in \figurename~\ref{fig:flids_vs_kits_masscan_flids}.

Unlike Mirai's C\&C behavior, Kitsune was able to detect the Merlin C\&C activity, which is noisier than Mirai's. Some packets related to the ICMP attack went undetected; however, all attacks included packets above the anomaly threshold: 0.9791 accuracy, 0.9891 F1 and 0.7392 MCC. Most anomalous packets from the scan-amplification data were also correctly classified: 0.9912 accuracy, 0.9955 F1 and 0.5988.

\subsubsection{non-clustered FL with weighted aggregation}

In this baseline, we train a single global model for all the clients, i.e., we are considering the $K=1$ case. The architecture for the anomaly detection autoencoder model is the same as in the clustered FL approach. We performed the federated hyperparameter tuning step, and the final optimizer selection is as follows: \textsc{ClientOpt} is Adam1 with $\eta = 0.005$, and \textsc{ServerOpt} is SGD with $\eta_s = 1.25$. The full FL training is performed with $E = 4$ and $R = 100$, as in the clustered case. Finally, we evaluate the trained global model on the mentioned validation-attack datasets. The anomaly detection threshold is selected in the same way as in the clustered version.

The evaluation of the global model on the devices that belonged to Cluster~0 and Cluster~2 resulted in subpar anomaly detection performance. All presented some normal packet instances with high reconstruction error that raised the anomaly detection threshold. The reconstruction error of anomalous samples was close to the error for normal traffic samples, yielding a near-zero value for F1 and MCC. While the anomaly detection performance could be improved by lowering the threshold at the expense of more false positives, most attacks would still be misclassified for all the tested validation-attack datasets.

On the contrary, the evaluation of the global model on the devices that belonged to Cluster 4 offered good anomaly detection performance, similar to the performance of the clustered anomaly detection version. While the anomaly detection threshold was larger than the clustered one due to some packets with higher reconstruction error, most anomalous packets were above the threshold. The measured metrics were greater than 0.9999 F1 or 0.9997 MCC for all the tested validation-attack datasets.

This baseline shows that the single FL global model is highly biased towards the six devices that belonged to Cluster~4 (IP cameras and stream consumers), which generate more data volume compared to the rest of the devices in the network. This effect might be caused because the aggregation server weights the client's contribution based on the number of training samples of that client. To adjust for this effect, the following FL baseline will equally weigh the contribution of all the federated clients.

\subsubsection{non-clustered FL without weighted aggregation}

We train a single global model ($K = 1$) using the same architecture for the anomaly detection autoencoder model as in the previous cases, but equally weighting the contribution of all the federated clients instead of by the amount of training data on each client. We use the following optimizers: \textsc{ClientOpt} is Adam1 with $\eta = 0.005$, and \textsc{ServerOpt} is SGD with $\eta_s = 1.0$. The full FL training is performed with $E = 4$ and $R = 100$, as in the previous cases.

The evaluation of the global model on the devices that belonged to Cluster~2 showed bad anomaly detection performance, similar to the previous baseline results with near zero metrics for F1 and MCC. However, the devices that belonged to Cluster~0 and Cluster~4 showed better anomaly detection metrics, but worse than the clustered FL approach: For the mirai-scan-load dataset, Cluster~0 showed 0.441 accuracy, 0.603 F1 and 0.111 MCC; Cluster~4 showed 0.947 accuracy, 0.599 F1, 0.636 MCC. For mirai-cnc-dos, Cluster~0 showed 0.213 accuracy, 0.194 F1 and 0.119 MCC; Cluster~4 showed 0.180 accuracy, 0.222 F1 and 0.094 MCC. For the merlin-cnc-dos dataset, Cluster~0 showed 0.146 accuracy, 0.205 F1 and 0.068 MCC; Cluster~4 showed 0.972 accuracy, 0.348 F1 and 0.452 MCC. Finally, for the masscan dataset, Cluster~0 showed 0.783 accuracy, 0.799 F1 and 0.642 MCC; Cluster~4 showed 0.999 accuracy, 0.816 F1 and 0.830 MCC.

\section{Conclusions} \label{sec:conclusions}

In this work, we have proposed a clustered FL architecture that allows training unsupervised anomaly and intrusion detection models in large networks of heterogeneous IoT devices. The proposed FL architecture does not need supervised data labeling, making it appropriate for real deployments where precise network traffic labeling is not feasible. To address the problems that arise with FL in heterogeneous environments, the proposed architecture includes an unsupervised device clustering algorithm that works by inspecting the parameters of the partially trained models. This clustering method is fully integrated into the FL training pipeline. It does not rely on any external fingerprinting tools or manual clustering methods, which can ease the implementation of FL-based architectures in deployment settings.

The architecture was implemented and evaluated on an emulated testbed comprised of multiple heterogeneous IoT and IIoT devices running real production libraries that generate traffic with a diverse set of network protocols. The proposed device clustering method showed successful grouping of the devices with similar communication patterns. However, as shown in the experiments, it must be noted that the clustering quality can be reduced in cases where the local training data in each device is not sufficient. Nevertheless, there was a wide margin of training data amount where the clustering results were mainly stable, and this can be mitigated by ensuring enough data is available before starting the process. It may also be advisable for the server to do several repetitions of the clustering step to ensure the stability of the process. Since the clustering step does not require much local training computation and only one round of communication is needed, it does not incur much cost. Additionally, training using FL exhibited a faster model convergence rate compared to the isolated edge method, especially for the devices that generate low volumes of training data.

The global models were evaluated on real attacks showing low false positive rates and high detection for most of the attacks. While few DoS-based attacks were not correctly classified as anomalous for some of the device clusters, the proposed model successfully detected stealthier malicious actions such as the Mirai C\&C heartbeat packets and slow scanning activities. In contrast, the comparison with the ML-based Kitsune network IDS showed that Kitsune correctly detected those DoS attacks but misclassified stealthier activity. This can indicate that for a more comprehensive detection, we could deploy alongside the clustered FL model a simpler model that, for instance, uses the frequency of packets over a time window to detect generic volumetric attacks. Additionally, the proposed clustered approach outperformed non-clustered FL baselines. Training a single global model for all the heterogeneous devices showed high bias towards the devices that generate more training data or a lack of generalization of the single global model. This highlights the advantage of personalization using clustered FL approaches for unsupervised network anomaly detection.

The IoT device types considered in the experimental scenario are devices with low mobility capabilities. The inclusion of devices with high mobility, such as intelligent vehicles and UAVs, presents additional challenges due to their frequent transitions between multiple wireless networks with varying quality of service. This movement can cause continuous changes in the extracted network features. Evaluating or adapting unsupervised clustered FL approaches in high mobility settings is a future line of work. If the data distribution of a device changes after the clustering process but before finishing the complete FL training, dynamic or soft clustering approaches might be considered to increase the flexibility when dealing with high-mobility IoT networks. Additionally, analyzing the root cause of an anomaly to distinguish intrusions or attacks between other causes, such as device updates, is another line of future work.

Lastly, we note that the unsupervised model training assumes that the devices are operating in normal conditions (i.e., during the training phase, the devices are not compromised). This assumption might not hold for some adversarial settings. Future work can include exploring how compromised or adversarial devices in the network affect the unsupervised device clustering stage of the proposed method. Compromised devices might deviate from other normal devices that should belong to the same cluster. This drift might be indicative of anomalous behavior, and the device can be flagged or filtered out before the FL process starts.

\section*{Acknowledgments}

The European commission financially supported this work through Horizon Europe program under the IDUNN project (grant agreement number 101021911). It was also partially supported by the Department of Economic Development, Sustainability and Environment of the Basque Government under the ELKARTEK 2023 program, project BEACON (with registration number 2023RTE00242510). Urko Zurutuza is part of the Intelligent Systems for Industrial Systems research group of Mondragon Unibertsitatea (IT1676-22), supported by the Department of Education, Universities and Research of the Basque Government.


\bibliographystyle{IEEEtran}
\bibliography{IEEEabrv,referencias,misreferencias} 


\vspace{1cm}

\textbf{Xabier~Sáez~de~Cámara} received his B.Sc. degree in Physics and Electronic Engineering from the Faculty of Science and Technology of the Basque Country University in 2015 and 2016, respectively. He holds an M.Sc. in Computational Engineering and Intelligent Systems from the University of the Basque Country. He is currently a Ph.D. student at IKERLAN in the Cybersecurity in Digital Platforms team and the Data Analysis and Cybersecurity research area at Mondragon Unibertsitatea, working on intrusion detection methods in IoT networks.\\

\textbf{Jose~Luis~Flores} is a researcher at Ikerlan Technology Research Center within the Cybersecurity in Embedded Systems team. He holds a M.Sc.~in Robotics and Advanced Control from the University of the Basque Country. His main interest is related to Artificial Intelligence and Cybersecurity. As such, the main lines he works on in each organization are Embedded System security at Ikerlan, and Machine Learning and Optimization at the university.\\

\textbf{Dr.~Cristóbal~Arellano} studied Computer Engineering at the University of the Basque Country, where he obtained his Ph.D. degree (with international mention) in Web Information Systems in 2013 (Cum Laude unanimously). He has been with IKERLAN since 2015 as a researcher and he currently is part of the Cybersecurity in Digital Platforms team. His current research interests include Cybersecurity in Cloud Platforms, Device Identity Management, DevSecOps, Federated Learning, Vulnerability Monitoring and Threat Detection. He has participated as an author or co-author in conferences such as WWW, ICWE, WISE, etc. He also has participated and led multiple European funded projects such as FP7 MONDO, UTEST, H2020 QUALITY and H2020 IDUNN.\\

\textbf{Dr.~Aitor~Urbieta} studied Computer Engineering at the University of Mondragon, where he obtained his Ph.D. degree (with international mention) in Computer Science in 2010 (Cum Laude unanimously). He has been with IKERLAN since 2007 where he currently leads the Cybersecurity in Digital Platforms research team. His current research interests include Cybersecurity in Digital Platforms, Internet of Things (IoT), Cybersecurity in Communication Protocols, Federated Learning, Blockchain, End-To-End Security, Vulnerability Monitoring, Threat Detection, Fog Computing, Edge Computing and IoT environment validation. He has participated as an author or co-author in more than 30 scientific publications in the previously mentioned areas, some of them Q1, published in national and international conferences and articles in JCR journals.\\

\textbf{Dr.~Urko~Zurutuza} is the principal investigator of the Intelligent Systems for Industrial Systems research group, and coordinator of the Data Analysis and Cybersecurity research area. He obtained his PhD in January 2008 at Mondragon Unibertsitatea, in collaboration with the Zürich IBM Research Lab. His research interests revolve around applications of Machine Learning to real world problems, and specially Cybersecurity. He has published more than 20 articles in high impact journals, more than 55 publications in blind peer-reviewed conferences, edited 3 books (2 of them as conference proceedings), and coauthored 7 book chapters. He is member of the Board of Directors of RENIC (National Network of Excellence in Cybersecurity Research), and serves in Steering Boards of leading international conferences such as DIMVA or RAID.

\end{document}